\documentclass[iop]{emulateapj}

\usepackage[colorlinks=true,linkcolor=blue,
  anchorcolor=blue,citecolor=blue,urlcolor=blue]{hyperref}  
\hypersetup{pdfauthor={Ho Seong HWANG}, 
  pdftitle={Testing weak-lensing maps of galaxy clusters}}

\slugcomment{Last updated: \today}
\shorttitle{Comparing dense galaxy cluster redshift surveys with weak lensing maps}
\shortauthors{Hwang et al.}

\usepackage{natbib}
\usepackage{subfigure}
\usepackage{url}

\begin{document}

\title{COMPARING DENSE GALAXY CLUSTER REDSHIFT SURVEYS WITH WEAK LENSING MAPS}



\author{Ho Seong Hwang\altaffilmark{1,2}, 
  Margaret J. Geller\altaffilmark{1}, 
  Antonaldo Diaferio\altaffilmark{3,4}, 
  Kenneth J. Rines\altaffilmark{5},
  H. Jabran Zahid\altaffilmark{1}}

\altaffiltext{1}{Smithsonian Astrophysical Observatory, 60 Garden Street, 
  Cambridge, MA 02138, USA; 
  hhwang@cfa.harvard.edu, mgeller@cfa.harvard.edu, harus.zahid@cfa.harvard.edu}
\altaffiltext{2}{School of Physics, Korea Institute for Advanced Study, 85 Hoegiro, 
  Dongdaemun-Gu, 130-722, Seoul, Korea}
\altaffiltext{3}{Dipartimento di Fisica, 
  Universit\`a degli Studi di Torino, V. Pietro Giuria 1, 10125 Torino, Italy; diaferio@ph.unito.it,}
\altaffiltext{4}{Istituto Nazionale di Fisica Nucleare (INFN), 
  Sezione di Torino, V. Pietro Giuria 1, 10125 Torino, Italy}
\altaffiltext{5}{Department of Physics and Astronomy, 
  Western Washington University, Bellingham, WA 98225, USA; kenneth.rines@wwu.edu}

\begin{abstract}

We use dense redshift surveys of nine galaxy clusters at $z\sim0.2$
  to compare the galaxy distribution in each system with
  the projected matter distribution from weak lensing.
By combining 2087 new MMT/Hectospec redshifts and 
  the data in the literature, we construct spectroscopic samples 
  within the region of weak-lensing maps of high (70--89\%) and uniform
  completeness.
With these dense redshift surveys,
  we construct galaxy number density maps using several galaxy subsamples.
The shape of the main cluster concentration 
  in the weak-lensing maps is similar to
  the global morphology of the number density maps 
  based on cluster members alone,
  mainly dominated by red members.
We cross correlate the galaxy number density maps with the weak-lensing maps.      
The cross correlation signal
  when we include foreground and background galaxies
  at 0.5$z_{\rm cl}<z<2z_{\rm cl}$
  is $10-23$\% larger than for cluster members alone
  at the cluster virial radius.
The excess can be as high as 30\% depending on the cluster.
Cross correlating the galaxy number density and weak-lensing maps
  suggests that superimposed structures close to the cluster
  in redshift space contribute more significantly
  to the excess cross correlation signal
  than unrelated large-scale structure along the line of sight.
Interestingly, 
  the weak-lensing mass profiles are not well constrained
  for the clusters with the largest cross correlation 
  signal excesses ($>$20\% for A383, A689 and A750).
The fractional excess in the cross correlation signal
  including foreground and background structures
  could be a useful proxy for assessing the reliability of 
  weak-lensing cluster mass estimates.
\end{abstract}


\section{INTRODUCTION}

Measurement of the mass distribution in galaxy clusters 
  is an important test of structure formation models
  \citep{duf08,pra12,rines13}.
Among the many measurements of the mass distribution of clusters,
 only weak lensing (e.g., \citealt{hoe07,ou08,oka10,ume14})
 and the caustic method based on galaxy kinematics \citep{dg97,dia99,serra11}
 reliably measure the cluster mass distribution
 regardless of the cluster dynamical state.
These measures can both extend into the infall region \citep{gel13}.
 
Weak-lensing has grown into
  a powerful probe of the distribution of dark matter 
  because it measures the total mass of a system directly
  regardless of the baryon content and/or dynamical state 
  \citep{clowe06,hut10,shan12}.
However, weak lensing includes the effect of structures
  projected along the line of sight \citep{hoe01};
  lensing provides a map of the total projected surface mass density.
There are also several unresolved systematic errors in weak-lensing analysis:
  e.g., systematic uncertainties
  in the measurements of gravitational shear and
  in the photometric redshift estimation of the distribution of lensed sources 
  \citep{hut13,uts14}.

The signal in weak-lensing maps centered on a cluster
  is generally dominated by the cluster itself.
However, there is an expected contribution to the signal from
 large-scale structure either associated or not associated with the cluster
 \citep{hoe01,hoe03,dod04}.
These structures are sometimes resolved in the weak-lensing maps,
 and can often appear even within the virial radii of clusters.
The lensing signal from these structures may introduce a bias and/or
  increase the uncertainty in 
  a cluster mass estimate based on lensing 
  \citep{hoe11,bk11,gru11,bahe12,coe12}.

Galaxy redshift surveys provide a map of the three-dimensional
  galaxy distribution. 
They can thus be used to resolve the structures
  along the line of sight 
  that may contribute to the total projected mass 
  \citep{hoe11cen,gel10,gel13,gel14f2}. 
Use of redshift surveys can also mitigate systematic errors
  resulting from the use of photometric redshifts 
  in weak-lensing analysis \citep{cou13}.
A direct comparison of the structures identified 
  in weak-lensing maps and in redshift surveys
  provides an important test of the issues limiting applications
  of weak lensing to measurement of cluster masses and
  mass profiles, and to the identification of galaxy clusters
  \citep{gel05,gel10,kurtz12,uts14,sta14}.

\begin{deluxetable*}{crrccccccc}
\tabletypesize{\scriptsize}
\tablewidth{0pc} 
\tablecaption{List of Galaxy Clusters
\label{tab-samp}}
\tablehead{
Name & R.A.$_{2000}$ & Decl.$_{2000}$ & z & Source of                  & Number of                   & Completeness & Subaru   & Radius of  & Number of  \\
     &      (deg)    &       (deg)    &   & redshifts\tablenotemark{a} & $z$ inside            & inside       & FOV      & entire field & of $z$ for \\
          &   &                         &                         &    & Subaru FOV\tablenotemark{b} &  Subaru FOV  & (arcmin) & (arcmin)    & entire field\tablenotemark{c} 
        }
\startdata
A383 &  42.01417 &  $-$3.52914 & 0.1887 & 1 & 411/153 & 82\%    & 20$^\prime\times$20$^\prime$ & 51 & 2544/275 \\
A267 &  28.17485 &  1.00709 & 0.2291 & 2 & 419/154 & 76\%       & 20$^\prime\times$20$^\prime$ & 33 & 1611/192 \\
A611 & 120.23675 & 36.05654 & 0.2880 & 3,4 & 335/129 & 76\%     & 20$^\prime\times$20$^\prime$ & 37 & 1836/295\\
A689 & 129.35600 & 14.98300 & 0.2789 & 2,5 & 333/119 & 73\%     & 20$^\prime\times$20$^\prime$ & 33 & 1282/220 \\
A697 & 130.73982 & 36.36646 & 0.2812 & 2,5,6 & 284/149 & 89\%   & 16$^\prime\times$16$^\prime$ & 33 & 1152/269\\
A750 & 137.24690 & 11.04440 & 0.1640 & 2,5 & 540/211 & 73\%     & 24$^\prime\times$24$^\prime$ & 33 & 1344/305\\
A963 & 154.26513 & 39.04705 & 0.2041 & 2,5,7 & 318/161 & 70\%   & 18$^\prime\times$18$^\prime$ & 33 & 1516/379\\
RXJ1720.1+2638 & 260.04183 & 26.62557 & 0.1604 & 2,8 & 220/121 & 89\% & 14$^\prime\times$14$^\prime$ & 33 & 1511/349\\
RXJ2129.6+0005 & 322.41647 & 0.08921 & 0.2339 & 2,9 & 156/71 & 71\%  & 12$^\prime\times$12$^\prime$ & 41 & 3522/249
\enddata
\tablenotetext{1}{1$-$\citet{gel14}, 2$-$\citet{rines13},  
3$-$Rines et al. 2014 (in preparation), 4$-$\citet{lem13}, 
5$-$This study, 6$-$\citet{gir06},
7$-$\citet{jaf13}, 8$-$\citet{owe11}, 9$-$\citet{dri10}.
We also add the redshifts from the SDSS DR10 and from NED.}
\tablenotetext{2}{Number of redshifts/Number of cluster members at $m_{\rm r,Petro,0}\leq 20.5$.}
\tablenotetext{3}{Number of redshifts/Number of cluster members regardless of magnitude range.}
\end{deluxetable*}

\begin{deluxetable*}{crccccrcc}
\tabletypesize{\footnotesize}
\tablewidth{0pc} 
\tablecaption{Redshifts in the fields of A689, A697, A750 and A963
\label{tab-gal}}
\tablehead{
Cluster & ID & SDSS ObjID & R.A.$_{2000}$ & Decl.$_{2000}$ & $m_{\rm r,Petro,0}$ & $z$ & $z$                        & Member\tablenotemark{c} \\
        &    & (DR10)     & (deg)         & (deg)          & (mag)               &     & Source\tablenotemark{b}    & 
}
\startdata
A689 &    1 &    1237667291574042910 & 128.795409 &  14.957459 &  18.939 & $ 0.00011\pm0.00003$ &  3 &  0 \\
A689 &    2 &    1237667538535055415 & 128.801168 &  15.053992 &  20.484 & $ 0.52506\pm0.00018$ &  3 &  0 \\
A689 &    3 &    1237667538535055402 & 128.802527 &  15.004511 &  17.237 & $ 0.00009\pm0.00001$ &  3 &  0 \\
A689 &    4 &    1237667538535055664 & 128.808520 &  15.046711 &  17.286 & $ 0.13889\pm0.00003$ &  3 &  0 \\
A689 &    5 &    1237667291574042897 & 128.812113 &  14.875166 &  17.978 & $ 0.00021\pm0.00001$ &  3 &  0 \\
A689 &    6 &    1237667538535055646 & 128.816686 &  15.004262 &  19.290 & $-0.00004\pm0.00005$ &  3 &  0 \\
A689 &    7 &    1237667291574042924 & 128.825021 &  14.912634 &  17.442 & $ 0.15485\pm0.00003$ &  3 &  0 \\
A689 &    8 &    1237667538535055723 & 128.827412 &  15.141970 &  17.646 & $ 0.15470\pm0.00004$ &  3 &  0 \\
A689 &    9 &    1237667538535055734 & 128.830726 &  15.171179 &  17.600 & $ 0.16997\pm0.00003$ &  3 &  0 \\
A689 &   10 &    1237667291574042958 & 128.839568 &  14.954218 &  17.674 & $ 0.00081\pm0.00001$ &  3 &  0
\enddata
\tablenotetext{1}{This table is available in its entirety in a machine-readable form in the online journal. A portion is shown here for guidance regarding its form and content.}
\tablenotetext{2}{(1) This study; (2) Rines et al. (2013); (3) SDSS DR10; (4) Girardi et al. (2006); (5) Jaff{\'e} et al. (2013); (6) NED.}
\tablenotetext{3}{(0) Cluster non-members; (1) Cluster members.}
\end{deluxetable*}

As an example of the impact of superimposed structure
  on a weak-lensing map, 
  \citet{gel14} used a deep, dense, nearly complete redshift survey 
  of the strong lensing cluster A383
  to compare the galaxy distribution with the weak-lensing results
  of \citet{oka10}.
The weak-lensing map of A383
  matches the galaxy number density map based on cluster members alone very well.
However, a secondary peak in the weak-lensing map
  is not clearly visible in the galaxy number density map 
  based on members (see their Fig. 8).
A galaxy number density map that includes foreground and background 
  galaxies around A383 produces a secondary peak
  consistent with the secondary weak lensing peak.
Thus, the secondary lensing peak apparently results from
  a superposition of foreground and background structures around A383.
The secondary peak lies within the virial radius of A383 and
  the cluster mass profile can be affected by superimposed structures 
  even within the virial radius.
The pilot study of A383 demonstrates the importance of dense
  redshift surveys to understand the projected mass distribution
  revealed by weak lensing.
  
Here, we compare maps based on dense redshift surveys
  with weak-lensing maps for nine additional clusters.
We compare the structures identified in the
  galaxy number density and weak-lensing maps
  by cross correlating the two.
With this sample we begin to quantify and elucidate the
  contribution to weak-lensing maps from structures
  superimposed along the line of sight.

Section \ref{data} describes the cluster sample and
 the data including the weak-lensing maps and the deep redshift surveys.
We measured 2087 new redshifts 
  to obtain galaxy number density maps 
  uniformly complete to $m_r=20.5$ for each cluster.
We compare galaxy number density maps of galaxy clusters with 
  the corresponding weak-lensing maps in Section \ref{results}.
We discuss the results and conclude
  in Sections \ref{discuss} and \ref{sum}, respectively.
Throughout,
  we adopt flat $\Lambda$CDM cosmological parameters:
  $H_0 = 100h$ km s$^{-1}$ Mpc$^{-1}$, 
  $\Omega_{\Lambda}=0.7$ and $\Omega_{m}=0.3$.

\section{DATA}\label{data}

Among the 30 clusters at $z\sim0.2$
  with high-quality Subaru weak-lensing maps in \citet{oka10},
  we select nine galaxy clusters with nearly complete redshift survey data.
With the inclusion of redshifts measured in this study,
  all nine clusters have an overall spectroscopic completeness
  within the region of weak-lensing map $>70\%$ for $m_{\rm r,Petro,0}\leq20.5$.

Table \ref{tab-samp} lists the nine clusters with 
  the redshift source, the number of redshifts and 
  the overall spectroscopic completeness within the weak-lensing maps,
  the field-of-views (FOVs) of the weak-lensing maps,
  the size of the entire cluster field where we compile the redshift data, and
  the number of redshifts in the entire field.
Here we describe the data we use
  to compare the redshift surveys with the weak lensing results.

\begin{figure*}
\center
\includegraphics[width=0.7\textwidth]{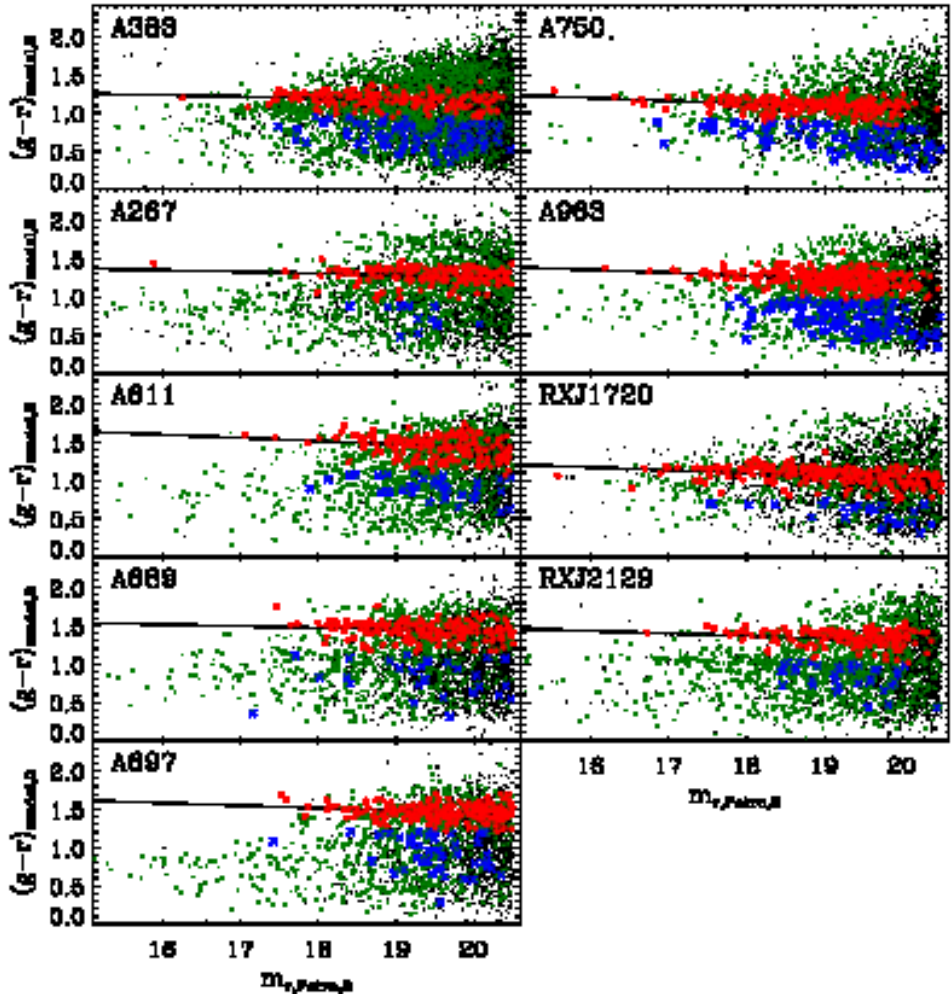}
\caption{Color-magnitude diagram for galaxies
  in the fields of nine galaxy clusters.
Black dots and green squares are galaxies without and with
  spectroscopic redshifts, respectively.
Red circles and blue crosses are red and blue member galaxies
  with spectroscopic redshifts, respectively.
Solid line is the best-fit of the red sequence in each cluster.
}\label{fig-cmr}
\end{figure*}

\subsection{Weak-Lensing Maps}

\citet{oka10} use Subaru/Suprime-Cam images \citep{miy02}
  for 30 X-ray luminous galaxy clusters
  at $0.15\leq z\leq0.3$ to determine the mass distribution around clusters.
Their paper includes a two-dimensional weak-lensing map of 
  each cluster field.
The map provides the normalized mass density field
  (i.e., the lensing convergence field, $\kappa$)
  relative to the 1$\sigma$ noise level expected
  from the intrinsic ellipticity noise. 
These maps are the basis for their derivation of 
  projected mass distribution around each cluster.

Typical FOVs of the maps are 
  $20^\prime\times20^\prime$. 
They smooth the map with the Gaussian, typical FWHM of $1^\prime.2$.
The range of FWHM is $1^\prime.0-1^\prime.7$.
We adopt their maps, listed in their Figures 16--45 
  (see their Appendix 3 for more details about the construction of the maps).
Among the 30 clusters in their sample, 
  we selected nine clusters where we measured new redshifts
  as necessary.

\subsection{The Cluster Redshift Surveys}\label{survey}

To compare the galaxy number density and weak-lensing map of a cluster,
  it is necessary to have a sufficiently dense, 
  nearly complete redshift survey of the cluster (e.g., \citealt{gel14}).
Among the 30 clusters at $z\sim0.2$
  with Subaru weak-lensing maps in \citet{oka10},
  we first select five clusters 
  with dense redshift data 
  in the Hectospec Cluster Survey (HeCS, \citealt{rines13}).
We supplement these data with redshifts from the literature
  \citep{gir06,dri10,owe11,lem13,jaf13,gel14},
  the Sloan Digital Sky Survey data release 10 (SDSS DR10, \citealt{ahn14}), and
  the NASA/IPAC Extragalactic Database (NED).

HeCS primarily observed galaxies close to the red sequence; 
  because our goal here is to study all line-of-sight structure 
  in the FOVs of the clusters, 
  we require additional observations to obtain 
  magnitude-limited redshift surveys.  
We made additional observations of four HeCS clusters 
  (A689, A697, A750 and A963) 
  in 2013 February and March with the 300 fiber Hectospec
  on the MMT 6.5m telescope \citep{fab05}.
The four clusters are within the footprint of the SDSS DR10.
To obtain a high, uniform spectroscopic completeness 
  at $m_{\rm r,Petro,0}\leq20.5$ within the field of
  the weak-lensing map,
  we weighted the spectroscopic targets
  according to the galaxy apparent magnitude independent of color.

\begin{figure*}
\center
\includegraphics[width=0.7\textwidth]{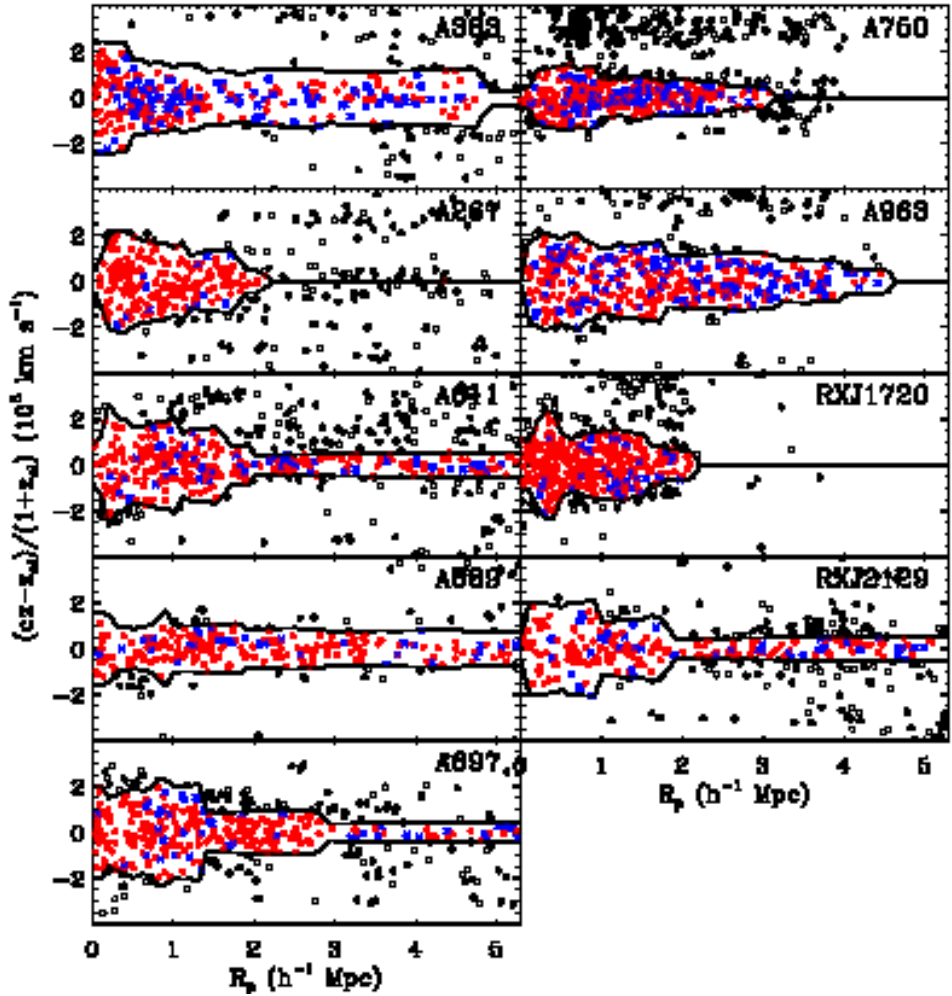}
\caption{Redshift (rest-frame clustercentric velocity) diagram
  for the nine clusters in this study.
Red circles and blue crosses are red and blue member galaxies, respectively.
Open circles are non-member galaxies.
Black lines are the caustics that distinguish member and non-member galaxies.
}\label{fig-caustic}
\end{figure*}

We used the 270 line mm$^{-1}$ grating of Hectospec
  that provides a dispersion of 1.2 \AA~pixel$^{-1}$ and 
  a resolution of $\sim$6 \AA.
We used 3$\times$20 minute exposures for each field, 
  and obtained spectra 
  covering the wavelength range $3650-9150$ \AA.
During the pipeline processing, 
  spectral fits are assigned a quality flag of ``Q'' 
  for high-quality redshifts, ``?'' for marginal cases, and ``X'' for poor fits.
We use only the spectra with reliable redshift measurement (i.e. ``Q'').
We set up two or three different Hectospec fields for each cluster, 
  and obtained 470--610 reliable redshifts per cluster.

Table \ref{tab-gal} lists the galaxy redshift data in
  the fields of the clusters.
We list 5294 galaxies with measured redshifts
  including 2087 new Hectospec redshifts
  for four clusters (A689, A697, A750 and A963).
The table contains the cluster name, identification, 
  SDSS DR10 ObjID, the right ascension (R.A.), declination (Decl.), 
  $r$-band Petrosian magnitude with Galactic extinction correction
  (from the SDSS DR10), the redshift ($z$) and its error, 
  the redshift source, and the cluster membership flag.

Figure \ref{fig-cmr} shows 
  Galactic extinction corrected (denoted by the subscript, ``0'') 
  $(g-r)_{\rm model,0}-m_{\rm r,Petro,0}$ 
  color-magnitude diagram for each cluster.
Black dots and green squares 
  are extended sources 
  without and with measured redshifts, respectively.
Red circles and blue crosses are red and blue cluster member galaxies.
Obviously, most bright galaxies have measured redshifts.
The plot includes all the sources in the entire cluster field 
  (see Table \ref{tab-samp} for the field size) 
  including the galaxies outside the weak-lensing maps,
  thus there are some bright galaxies without measured redshifts.
We explain the details of the spectroscopic completeness 
  in the region covered by each weak-lensing map in the next section.

\begin{figure*}
\center
\includegraphics[width=0.7\textwidth]{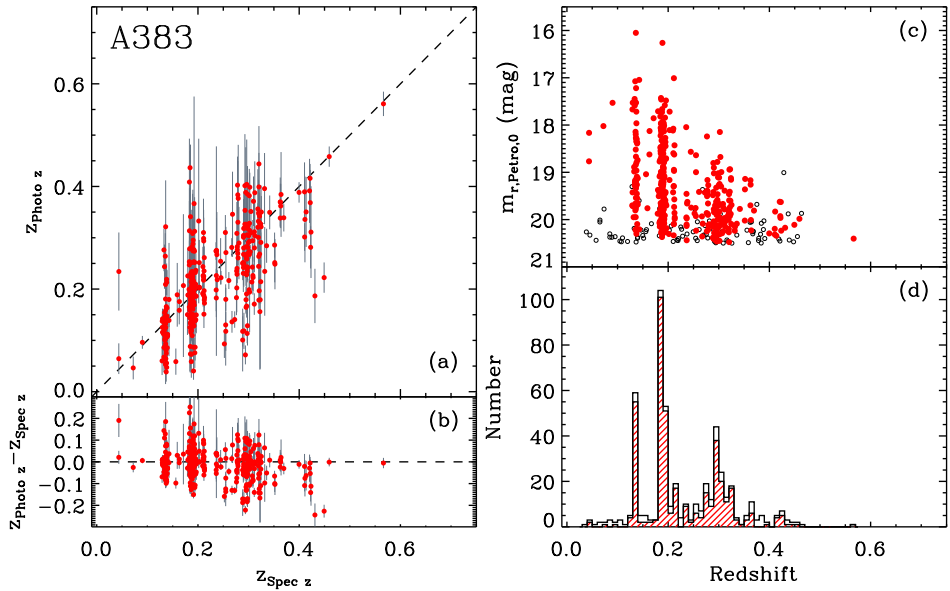}
\caption{({\it a-b}) Comparison between photometric redshifts
  \citep{csa03} and spectroscopic redshifts
  for the galaxies with $m_{\rm r,Petro,0}<20.5$
  inside the field of weak-lensing map (see Fig. \ref{fig-a383spat}).
  (c) $r$-band Petrosian magnitudes of galaxies
  with spectroscopic (red filled circles) and photometric (open circles)
  redshifts as a function of redshift.
(d) Redshift distribution of galaxies
  with spectroscopic (red hatched histogram) and photometric (open histogram)
  redshifts.
}\label{fig-a383photz}
\end{figure*}

To determine the membership of galaxies in each cluster,
  we use the caustic technique \citep{dg97,dia99,serra11}, 
  originally devised to
  determine the mass profiles of galaxy clusters.
The technique produces a useful tool for determining cluster membership.
Analysis of galaxy clusters in a cosmological $N$-body simulation
  indicates that the caustic technique identifies
  true cluster members with $95$\% completeness within 3$r_{200}$.
The contamination of interlopers in the member galaxy catalog
  is only 2--8\% at 1--3$r_{200}$ \citep{ser13}.

The caustic technique first uses the redshifts and the position on the sky
  of the galaxies to determine a hierarchical center of the cluster
  based on a binary tree analysis.
We then plot the rest-frame clustercentric velocities of galaxies
  as a function of projected clustercentric radius
  centered on the hierarchical center.
Figure \ref{fig-caustic} shows this phase-space diagram
  for each cluster; the expected trumpet-shaped pattern is obvious
  \citep{kai87,rg89}.
The caustics (solid lines) generally agree with the lines based on a visual impression.
They are cleanly defined especially at small radii.    
We use all the galaxies with measured redshifts regardless of
  their magnitudes for determining the membership, 
  but restrict our analysis to the galaxies at
  $m_{\rm r,Petro,0}\leq20.5$ within the weak-lensing maps
  for comparison between the galaxy number density and weak-lensing maps.

To segregate the red and blue cluster populations,
  we define a red sequence from a linear fit 
  to the bright, red member galaxies in Figure \ref{fig-cmr}
  (e.g., $m_{\rm r,Petro,0}<19.5$ and $1<(g-r)_{\rm model,0}<2$ for A383).
The solid line in each panel shows the red sequence for each cluster.
The typical rms scatter ($\sigma$) around the red sequence is $\sim0.1$ mag.
A line $3\sigma$ blueward of the red sequence separates
  the red and blue members.

When we construct galaxy number density maps in Section \ref{numden},
  we also use SDSS photometric redshifts for galaxies 
  without spectroscopic redshifts \citep{csa03}.
The rms uncertainties of photometric redshifts
  for the galaxies at $m_r<18$ and $m_r<21$ are 
  $\sim$0.035 and $\sim$0.103, respectively \citep{csa03}.
The redshift range necessary for
  constructing galaxy number density maps to be compared with weak-lensing maps,
  is broad enough not to be significantly affected by these uncertainties.

As a prototypical example,
 we compare spectroscopic and photometric redshifts
 for galaxies in the field of A383
 in the left panel of Figure \ref{fig-a383photz}.
As expected, the photometric redshifts roughly agree 
  with the spectroscopic redshifts with a large scatter.
The top right panel shows
  $r$-band Petrosian magnitudes of galaxies with
  spectroscopic (red filled circles) and photometric (open circles)
  redshifts as a function of redshift.
The bottom right panel shows 
  the redshift histogram for each galaxy sample
  (red hatched and open histograms 
  for spectroscopic and photometric redshifts, respectively).
The foreground and background structures 
  at $z\sim0.14$ and $z\sim0.3$, respectively ($z_{\rm A383}=0.1887$),
  are apparent in the histogram. 
As expected, the photometric redshifts contribute little;
  they are important mainly for faint galaxies.
The comparison of spectroscopic and photometric redshifts
  for the other eight clusters is similar to A383, and we thus
  do not display them.

\section{Results}\label{results}

Here we construct galaxy number density maps 
  using several galaxy subsamples in Section \ref{numden}, 
  and cross correlate them with the weak-lensing maps in Section \ref{cross}.
We first test whether red cluster members can reproduce
  the main weak-lensing peak (see e.g., \citealt{zit10}), and then add
  other populations successively 
  (i.e. foreground/background galaxies along the line of sight)
  to gauge their impact on the weak-lensing maps.

\subsection{Galaxy Number Density Maps}\label{numden}

To construct galaxy number density maps based on
  a spectroscopic sample of galaxies,
  it is important to understand any bias 
  introduced by the spectroscopic observations,
  especially spectroscopic incompleteness.
The pixel size and smoothing scale we need are set by
  the weak lensing maps of \citet{oka10}:
  e.g., 201$\times$201 pixels for the
  $20^\prime\times 20^\prime$ weak-lensing map and
  Gaussian smoothing scale of FWHM $=1\arcmin.2$ for A383.

\begin{figure*}
\center
\includegraphics[width=0.8\textwidth]{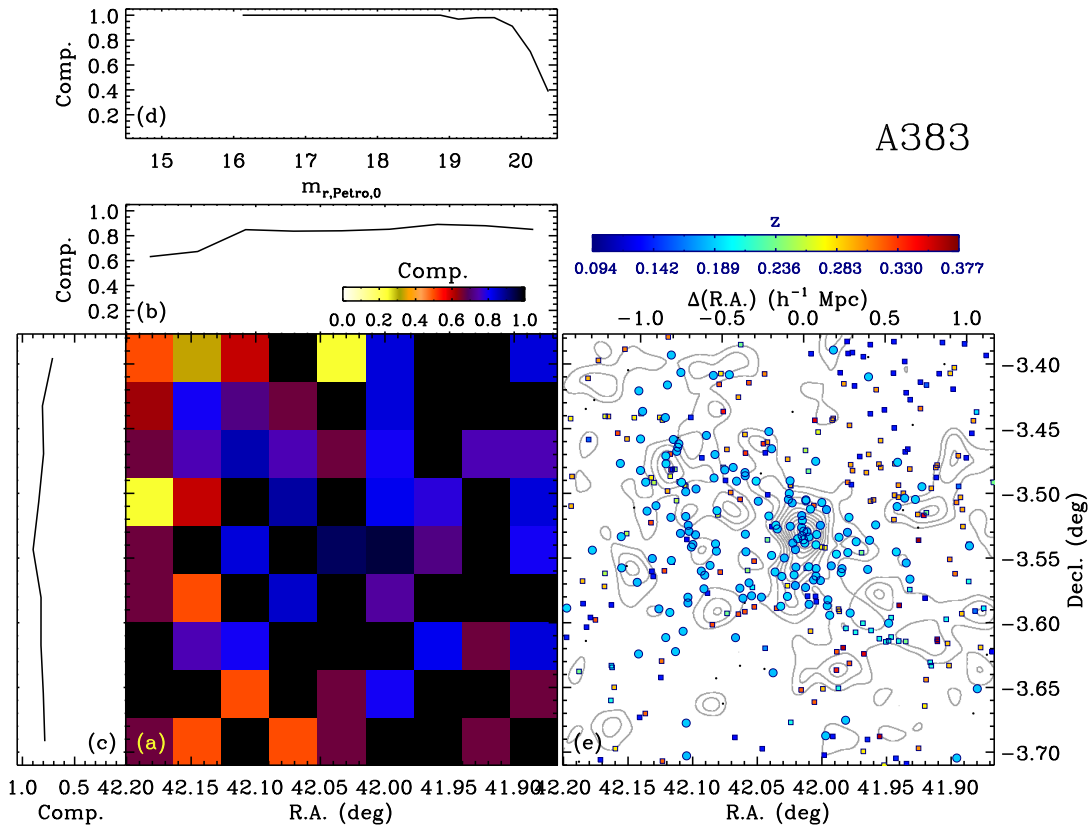}
\caption{(a) Two-dimensional spectroscopic completeness
  in the field of A383 
  as a function of right ascension and of declination.
Integrated spectroscopic completeness as a function of right ascension (b), 
  declination (c), and $r$-band magnitude (d).
(e) Spatial distribution of galaxies at $m_{\rm r,Petro,0}\leq20.5$ 
  (open circles: members, squares: non-members 
  inside the color bar redshift window, dots: non-members outside the 
  color bar redshift window).
Gray contours are the weak-lensing map of \citet{oka10}.
}\label{fig-a383spat}
\end{figure*}

\subsubsection{Spectroscopic Completeness}

The left panel (a) of Figure \ref{fig-a383spat}
  shows a two-dimensional map of the spectroscopic completeness
  for $m_{\rm r,Petro,0}\leq20.5$
  as a function of right ascension and declination, 
  matched to the FOV of the weak-lensing map of A383.
The two-dimensional completeness map is in 10$\times$10 pixels 
  for the $20^\prime\times 20^\prime$ weak-lensing map.
Panels (b-c) show the integrated completeness
  as a function of right ascension and declination, respectively.
The overall completeness in this field is $\sim$81\%.
Although there are three pixels with a completeness $<50\%$, 
  the completeness changes little with right ascension and declination.
Panel (d) shows the integrated completeness
  as a function of $r$-band magnitude;
  it drops only at $m_{\rm r,Petro,0}>20$.

We also show the spatial distribution of galaxies
 superimposed on the A383 weak-lensing map of \citet{oka10} 
 in the right panel.
Many member galaxies of A383 (open circles) are distributed
  around the peaks of the weak-lensing map,
  but some foreground and background galaxies (squares) 
  are located around the peaks.
The plots for other clusters are in the Appendix
 (see Figs. \ref{fig-spat2}--\ref{fig-spat5}).

\begin{figure*}
\center
\includegraphics[width=5.7in]{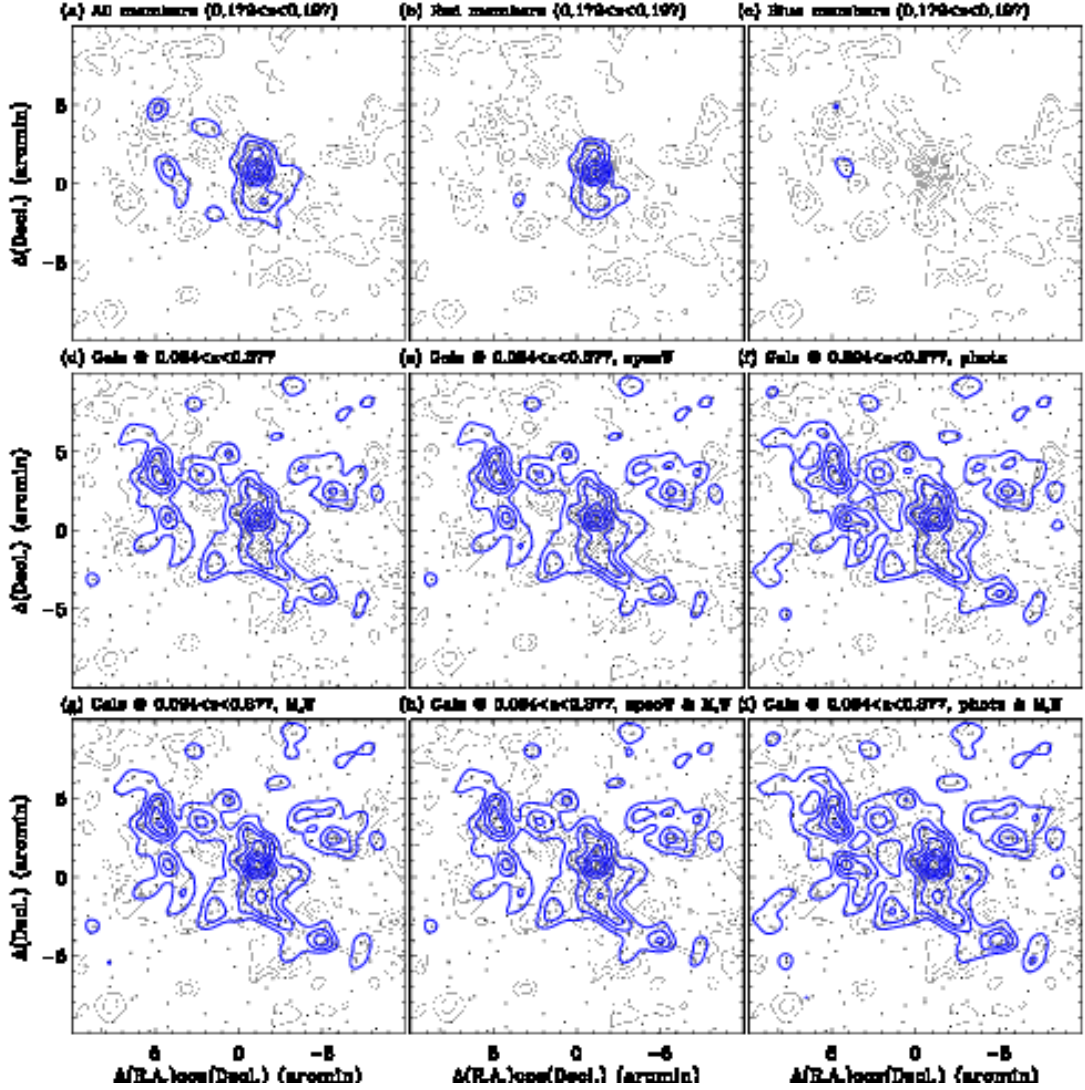}
\caption{A383 galaxy number density maps 
  for several subsamples at $m_{\rm r,Petro,0}<20.5$.
The lowest cluster surface number density contours is 
  1.61 galaxies arcmin$^{-2}$ and 
  the contours increase in steps of 0.85 galaxy arcmin$^{-2}$.
We smooth the contours with the Gaussian of FWHM $=1\arcmin.2$,
  the same smoothing scale as for weak-lensing maps of \citet{oka10}.
Top row is for (a) all cluster members, 
  (b) red cluster members, and (c) blue cluster members.
Middle row is for the galaxies 
 in the redshift range $0.5z_{\rm cl}\leq z\leq2z_{\rm cl}$
 (i.e., $0.094<z<0.377$ for A383):
  (d) without any weights,
  (e) weighted by spectroscopic completeness, and
  (f) complemented by galaxies with photometric redshifts.
Bottom row is for the same sample as for in the middle row,
  but for weighted by stellar masses of galaxies as well:
  (g) without any weights,
  (h) weighted by spectroscopic completeness, and
  (i) complemented by galaxies with photometric redshifts.
}\label{fig-a383numden}
\end{figure*}

\subsubsection{Construction of Galaxy Number Density Maps: 
 Revisiting A383}\label{densub}

To study the relative contribution of cluster and foreground/background
  structure to the weak-lensing maps,
  we construct an extensive set of
  galaxy number density maps based on
  several galaxy subsamples.
We smooth the contours with the same Gaussian FWHM
  as for the weak-lensing map of \citet{oka10}: 
  e.g., FWHM$=1\arcmin.2$ for A383.
We also use the same pixel size as for the weak lensing map
  (e.g., 201$\times$201 pixels for 
  $20^\prime\times 20^\prime$ FOV of A383).   
Figure \ref{fig-a383numden} shows these number density maps
 (blue contours)
 superimposed on the weak-lensing map (gray contours).
The top panels are based on 
  cluster member galaxies alone (all, red and blue members from left to right).
We separate red and blue galaxies based on 
  their positions in the $(g-r)_{\rm model,0}-m_{\rm r,Petro,0}$ color-magnitude
  diagram (see Section \ref{survey}).

As in \citet{gel14},
  the global morphology of the spatial distribution
  of cluster members alone (red and red plus blue populations) is similar to 
  the shape of the main concentration in the weak-lensing map;
  both show a north-south elongation (see Panel a).
The density map based on red members (Panel b) is also similar to
  the cluster shape in the weak-lensing map.
Interestingly, blue members alone have little correspondence 
  with the weak-lensing map (Panel c).
This comparison supports the idea that 
  the red population provides a reasonable tracer
  of the mass distribution of a galaxy cluster
  provided that the selection is dense and broad enough 
  around the red sequence \citep{rines13,gel14}.
The comparison also suggests that the assumption that 
  red cluster galaxies trace the dark matter distribution
  is a very reasonable approach for strong lensing models
  \citep{bro05,zit09,med10}.

The middle panels show contours for galaxies 
  in the redshift range $0.5z_{\rm cl}\leq z\leq2z_{\rm cl}$
 (i.e., $0.094\leq z \leq0.377$ for A383)
 where the contribution to the lensing signal may be significant \citep{hu99}.
The number density map in Panel (d) is based on galaxies 
  with measured redshifts.
To account for galaxies without measured redshifts,
 we use two methods to construct the maps;
 we correct statistically for spectroscopic incompleteness
 and we use photometric redshifts.
To correct for spectroscopic incompleteness,
  we compute the spectroscopic completeness
  for each object with a measured redshift
  in a three-dimensional parameter space:
  $r-$band magnitude, right ascension and declination.
The right panels in Figure \ref{fig-a383spat} show
  the spectroscopic completeness
  in this parameter space.
We thus weight each galaxy by the inverse of 
  the spectroscopic completeness
  to derive the galaxy number density map, and 
  show it in Panel (e).
We also use SDSS photometric redshifts
  for the galaxies without spectroscopic redshifts 
  (see Section \ref{survey} for details).
The corresponding number density map is in Panel (f).

In the bottom panels,
  we use the same galaxy samples as in middle panels,
  but additionally we weight each galaxy with the stellar mass.
We compute stellar masses 
  using the SDSS five-band photometric data with the 
  Le Phare\footnote{\url{http://www.cfht.hawaii.edu/~arnouts/lephare.html}} 
  code \citep{arn99,ilb06}.
Details of the stellar mass estimates are in \citet{zah14}.
Three panels show the maps
  based on galaxy samples of observed galaxies (g),
  weighted by spectroscopic completeness (h), and
  supplemented by photometric redshifts (i).

It is interesting that 
  the global morphology of the number density maps in Panels (d--i) are similar
  despite the different correction methods.
No matter how we weight the data,
  the correspondence between the galaxy number density 
  and weak-lensing maps
  is more remarkable when we include foreground and background galaxies
  in the galaxy number density maps (Panels d--i),
  underscoring the contribution of large-scale structure
  along the line of sight to the weak-lensing signal.

\begin{figure*}
\center
\includegraphics[width=0.7\textwidth]{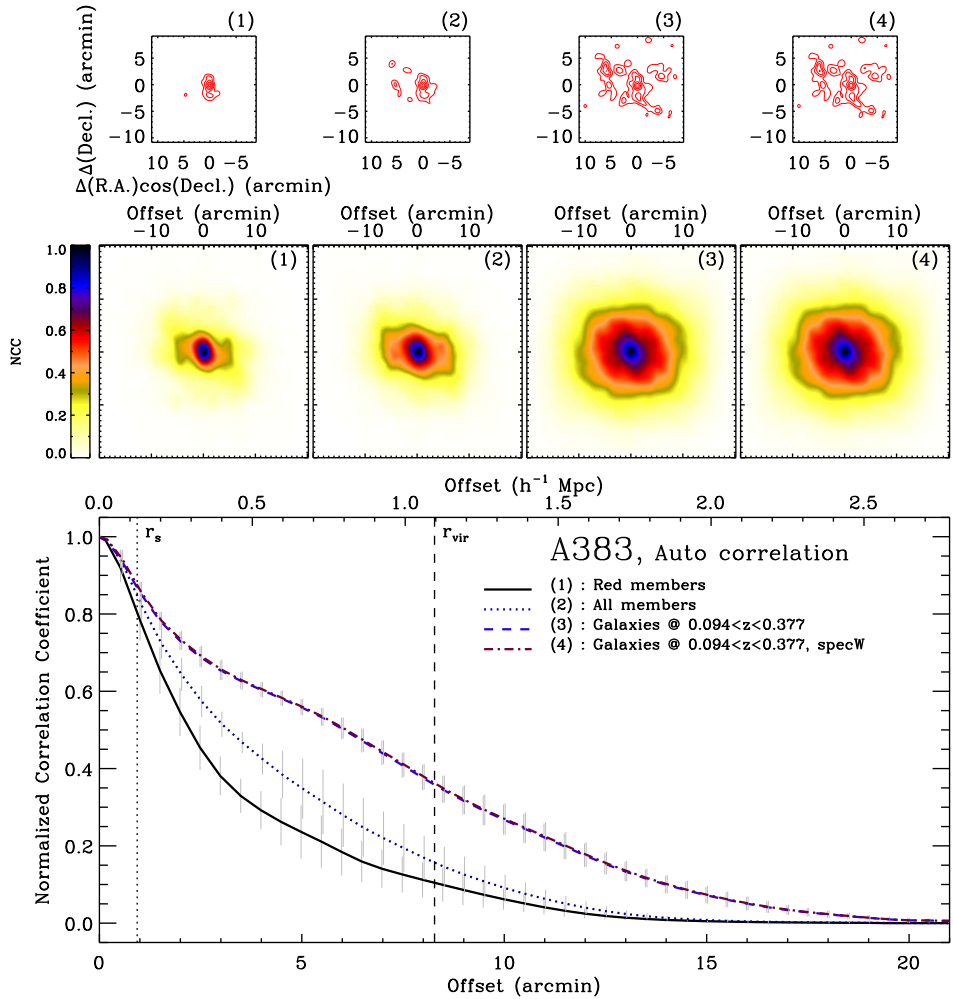}
\caption{Normalized auto correlation of A383 galaxy number density maps.
({\it Top}) Red contours indicate galaxy number density maps for four subsamples.
({\it Middle}) Two-dimensional normalized auto correlation maps.
({\it Bottom}) Azimuthally averaged correlation signal
  as a function of offset.
Vertical dotted and dashed lines indicate
  $r_{s}$ and $r_{\rm vir}$ from \citet{oka10}, respectively.
}\label{fig-afcc}
\end{figure*}

\subsection{Cross Correlation between Galaxy Number Density and 
  Weak-lensing Maps}\label{cross}

Here, we cross correlate the weak-lensing and 
  galaxy number density maps
  for several of the galaxy subsamples.
We use the normalized cross correlation (NCC), 
  widely used in image processing \citep{gw02}.
It is defined by
\begin{eqnarray}\label{eq-ncc}
NCC(x,y)  & = & \frac{\sum\limits_{i,j} I_1(i,j) I_2(i+x,j+y)}
  {\sqrt{\sum\limits_{i,j} I_1^2(i,j) } 
   \sqrt{\sum\limits_{i,j} I_2^2(i+x,j+y)}}
\end{eqnarray}
where $I_1(i,j)$ and $I_2(i,j)$ are pixel values
  of the galaxy number density and weak-lensing maps, respectively.

\subsubsection{A383}\label{a383}

To test our code and to understand what we can learn from the cross correlation, 
  we first calculate the auto correlation for a galaxy number density map
  for a simulated cluster.
We construct a mock galaxy catalog with 1000 galaxies in a cluster
  following the Navarro-Frenk-White profile 
  (NFW; \citealt{nfw97}).
We use the NFW profile of A383 with 
  $r_{200}=1.184$ ($h^{-1}$Mpc) and $c_{200}=6.51$
  derived in \citet{new13}, 
  and we show the results in the Appendix 
  (see Figs. \ref{fig-simnum} and \ref{fig-simcor}).
As expected, the normalized auto correlation signal is equal to 
  unity at zero offset, and 
  the correlation signal decreases with offset.
It converges to zero at large radius (i.e. $r>r_{\rm 200}$);
  in other words, there is intrinsically no auto correlation signal at this radius
  and/or the auto correlation signal becomes small 
  because at the edge of the maps.

\begin{figure*}
\center
\includegraphics[width=0.7\textwidth]{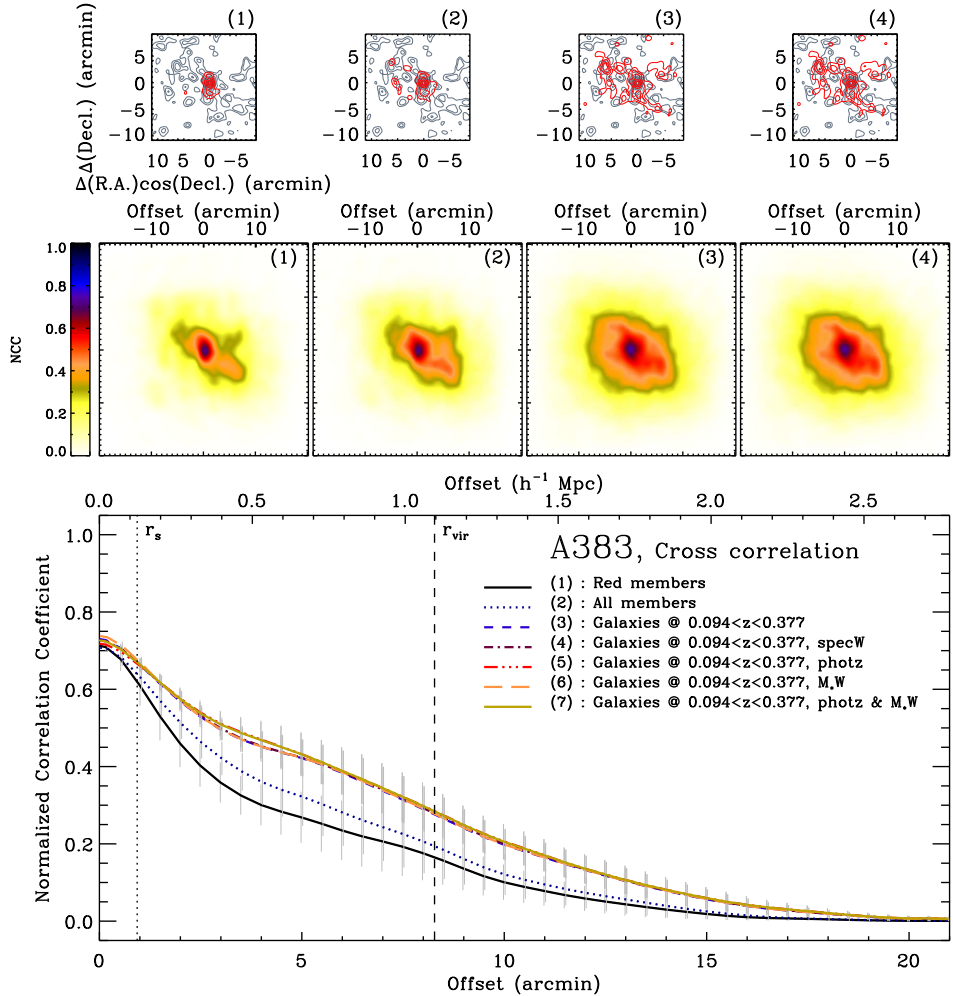}
\caption{Normalized cross correlation of A383 galaxy number density maps.
({\it Top}) Gray and red contours indicate weak-lensing and 
  galaxy number density maps, respectively.
We use seven galaxy subsamples listed in the bottom panel, but
  show only four examples. 
({\it Middle}) Two-dimensional normalized cross correlation maps.
({\it Bottom}) Azimuthally averaged correlation signal
  as a function of offset.
Vertical dotted and dashed lines indicate
  $r_{s}$ and $r_{\rm vir}$ from \citet{oka10}, respectively.
}\label{fig-cfcc}
\end{figure*}

We also calculate the auto correlation for galaxy number density maps
  representing several galaxy subsamples of A383,
  and show the results in Figure \ref{fig-afcc}.
The top panels show the galaxy number density maps (red contours)
  for four subsamples:
  (1) red members, (2) all members, 
  (3) galaxies at $0.5z_{\rm cl}\leq z\leq2z_{\rm cl}$, and
  (4) galaxies at $0.5z_{\rm cl}\leq z\leq2z_{\rm cl}$ weighted by their
  spectroscopic completeness (see Figure \ref{fig-a383spat}).  
The middle panels show the two-dimensional auto correlation map for each case.
The bottom panel shows the azimuthally averaged auto correlation signal
  derived from each of the middle panel.
The gray error bar indicates the dispersion in
  the two-dimensional correlation signal at each offset.
As expected, the normalized correlation signal is
  unity at zero offset.
The auto correlation signal converges to zero at $r>r_{\rm vir}$
  for cases based on members alone (solid and dotted lines).
However, the signal for cases including
  the galaxies at $0.5z_{\rm cl}\leq z\leq2z_{\rm cl}$
  (dashed and dot-dashed lines)
  is not negligible even at $r\sim r_{\rm vir}$, and 
  decreases slowly; this effect results mainly 
  from foreground and background galaxies along the line of sight
  that contribute to the auto correlation signal.

Figure \ref{fig-cfcc} shows the results of 
 cross correlating the galaxy number density maps 
 and weak-lensing map.
As in Figure \ref{fig-afcc},
  the top panels show the galaxy number density maps (red contours)
  for four galaxy subsamples superimposed 
  on the weak-lensing maps (gray contours):
  (1) red members, (2) all members, 
  (3) galaxies at $0.5z_{\rm cl}\leq z\leq2z_{\rm cl}$, and
  (4) galaxies at $0.5z_{\rm cl}\leq z\leq2z_{\rm cl}$ weighted by their
  spectroscopic completeness (see Figure \ref{fig-a383spat}).  
The middle panels show the two-dimensional cross correlation map for each case.
We use seven galaxy subsamples in the cross correlation, 
  but show only four cases
  in the top and middle panels for ease of view.
The bottom panel shows the azimuthally averaged cross correlation signal
  for the seven galaxy subsamples.

Case (1) based on red member galaxies gives, not surprisingly,
  the narrowest correlation peak.
The signal converges to zero at $r>r_{\rm vir}$
  for the cases based on member galaxies alone (solid and dotted lines).
The cases including
  galaxies at $0.5z_{\rm cl}\leq z\leq2z_{\rm cl}$ (i.e., cases 3--7)
  are similar to one another, and
  show a broader distribution.
The cross correlation signal is always larger than 
  for the cases based on member galaxies alone, 
  indicating the non-negligible contribution of foreground and background
  galaxies to the weak-lensing map.

\begin{figure*}
\center
\begin{tabular}{c}
\includegraphics[width=0.8\textwidth]{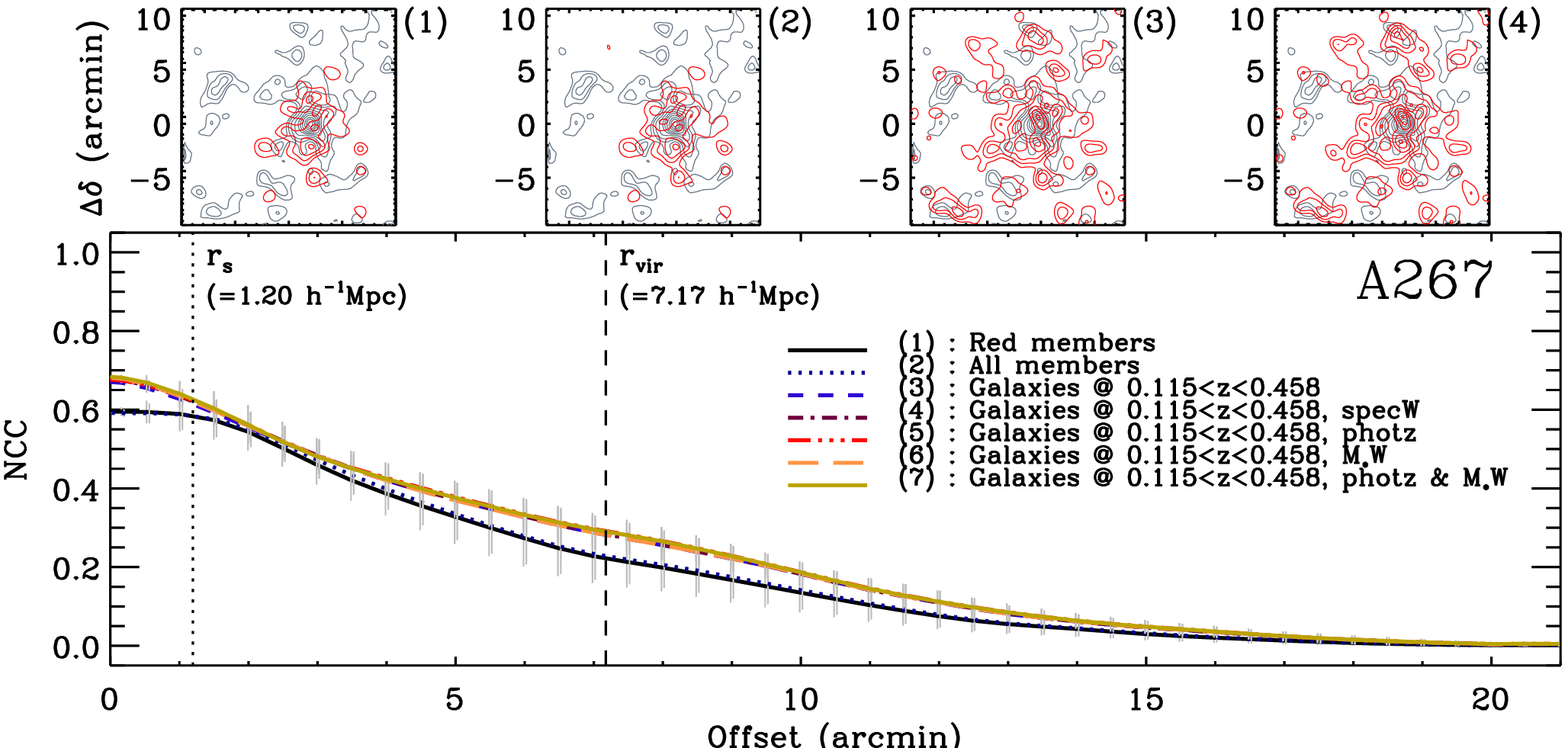} \\
\includegraphics[width=0.8\textwidth]{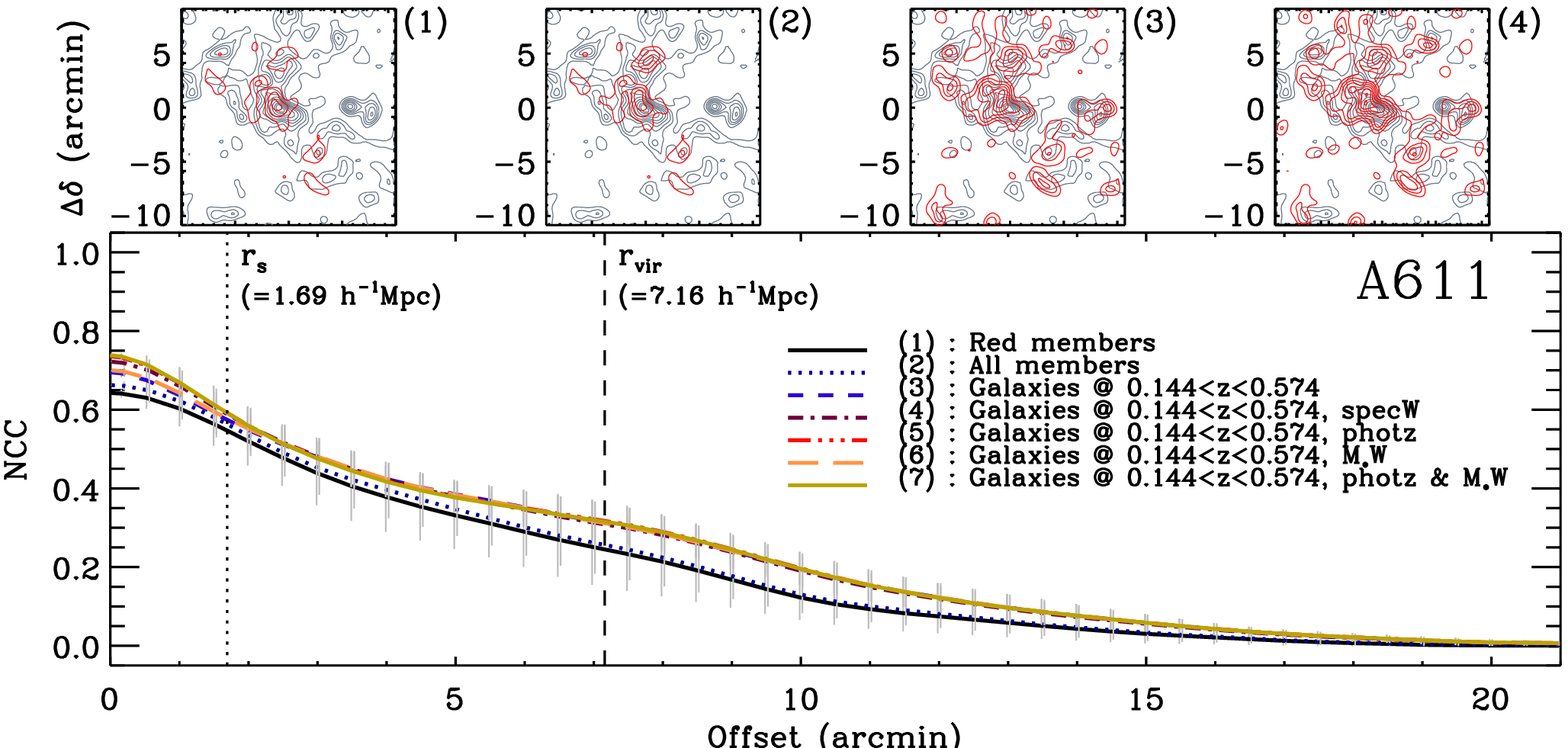} \\
\end{tabular}
\caption{Same as Fig. \ref{fig-cfcc} (except for 2D NCC map),
 but for A267 and A611.
Gray and red contours indicate weak-lensing and 
  galaxy number density maps, respectively.
}\label{fig-cfcc2}
\end{figure*}

\subsubsection{Comments on Individual Clusters}\label{comments}

\begin{figure*}
\center
\begin{tabular}{c}
\includegraphics[width=0.8\textwidth]{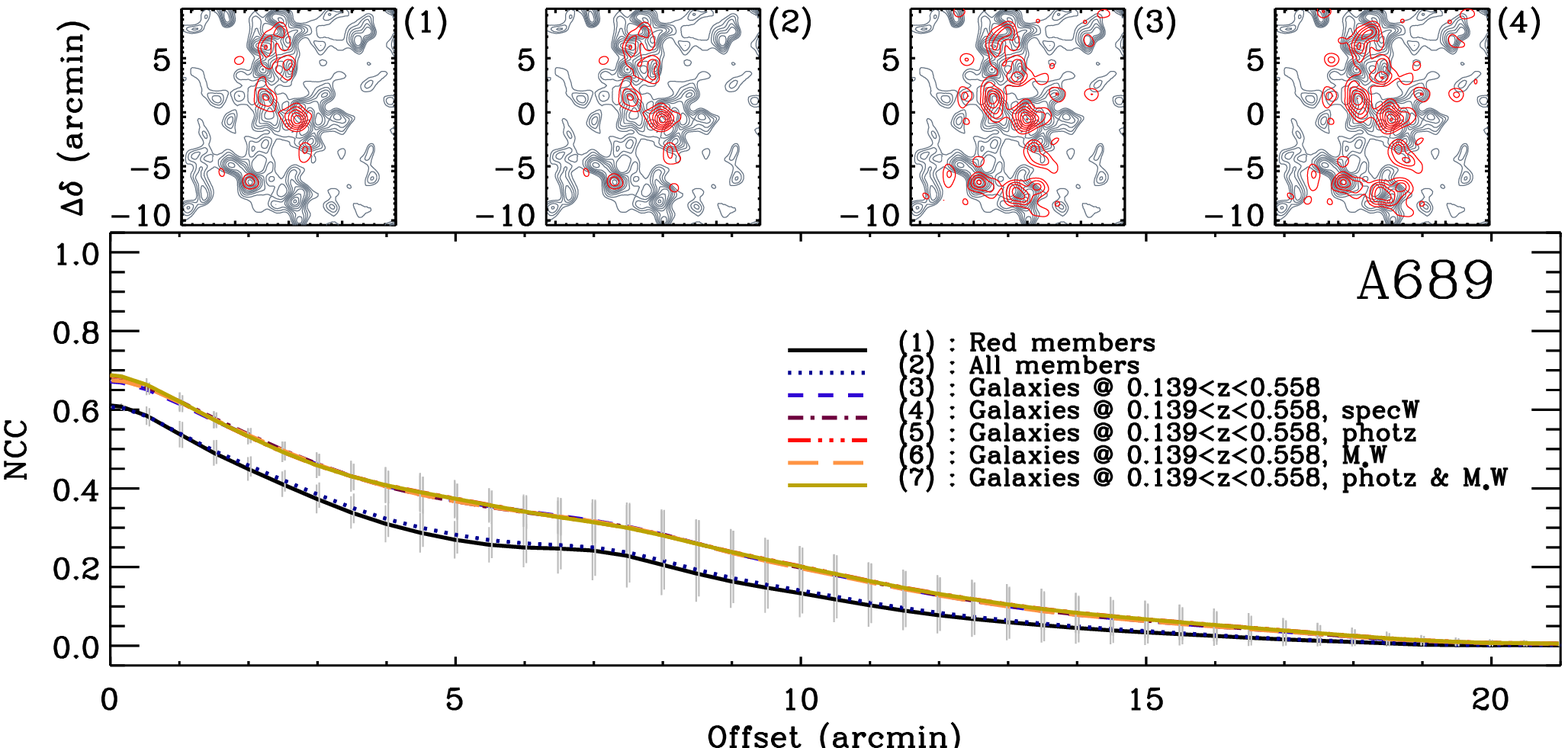} \\
\includegraphics[width=0.8\textwidth]{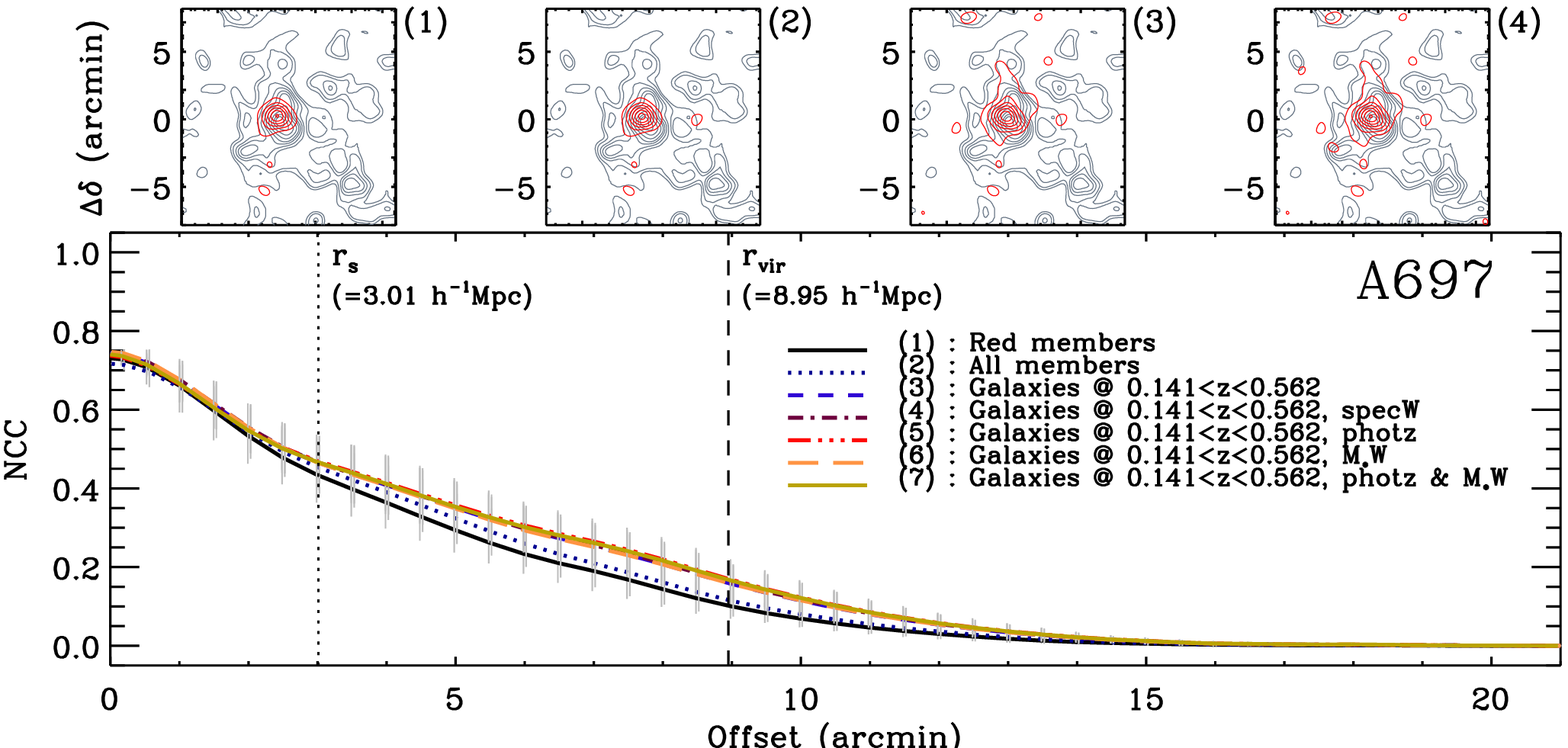} \\
\end{tabular}
\caption{Same as Fig. \ref{fig-cfcc} (except for 2D NCC map),
 but for A689 and A697.
Gray and red contours indicate weak-lensing and 
  galaxy number density maps, respectively.
}\label{fig-cfcc3}
\end{figure*}

\begin{figure*}
\center
\begin{tabular}{c}
\includegraphics[width=0.8\textwidth]{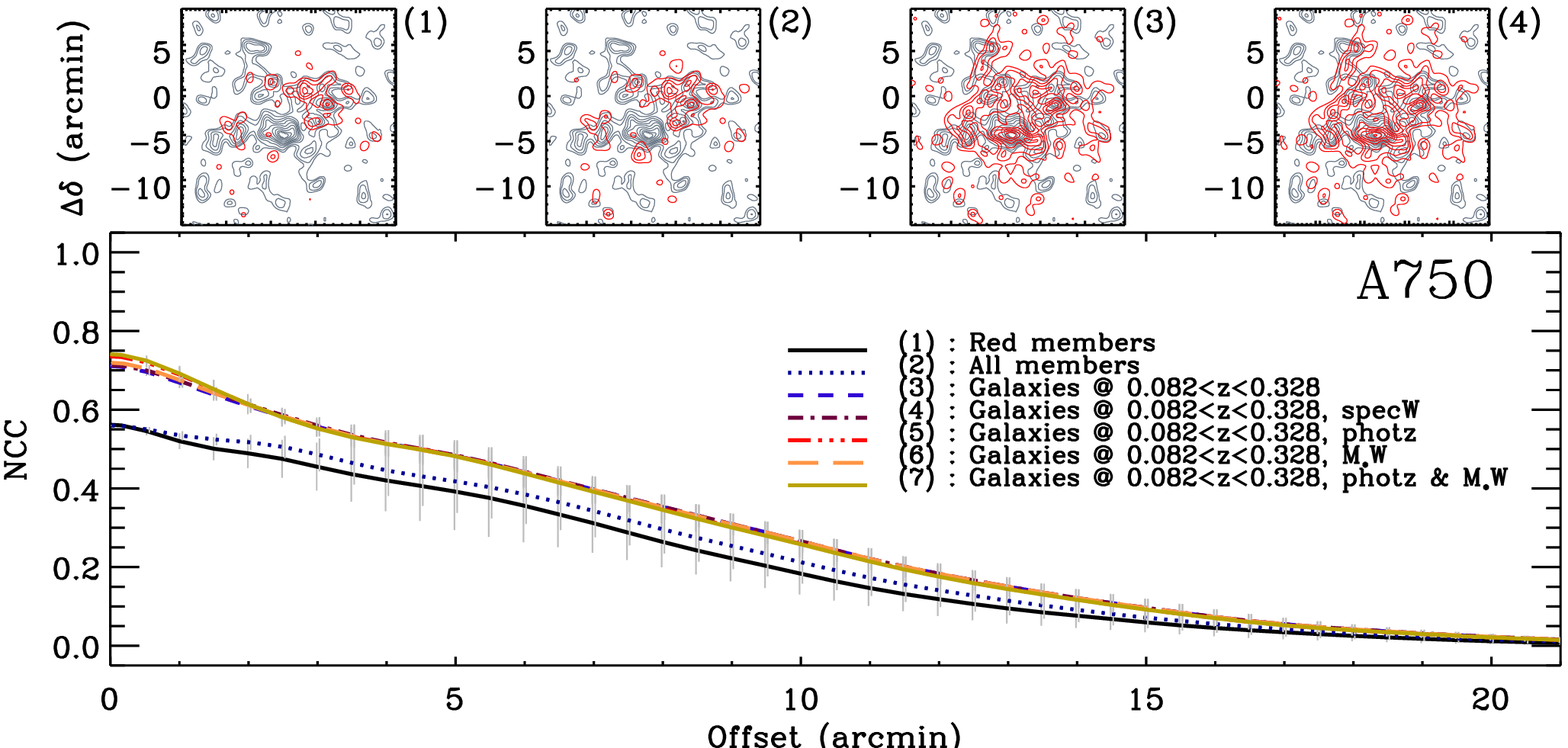} \\
\includegraphics[width=0.8\textwidth]{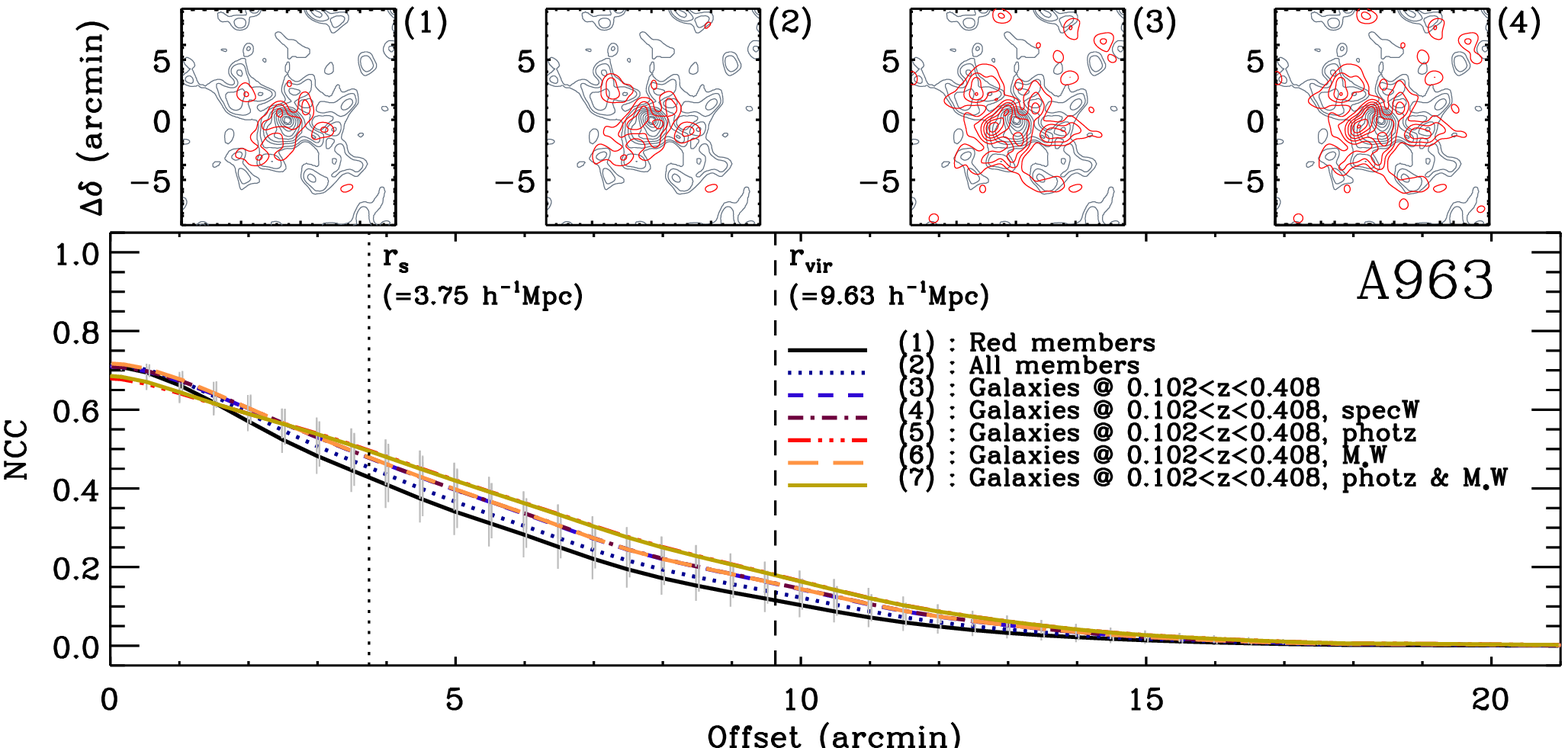} \\
\end{tabular}
\caption{Same as Fig. \ref{fig-cfcc} (except for 2D NCC map),
 but for A750 and A963. 
Gray and red contours indicate weak-lensing and 
  galaxy number density maps, respectively.
}\label{fig-cfcc4}
\end{figure*}

\begin{figure*}
\center
\begin{tabular}{c}
\includegraphics[width=0.8\textwidth]{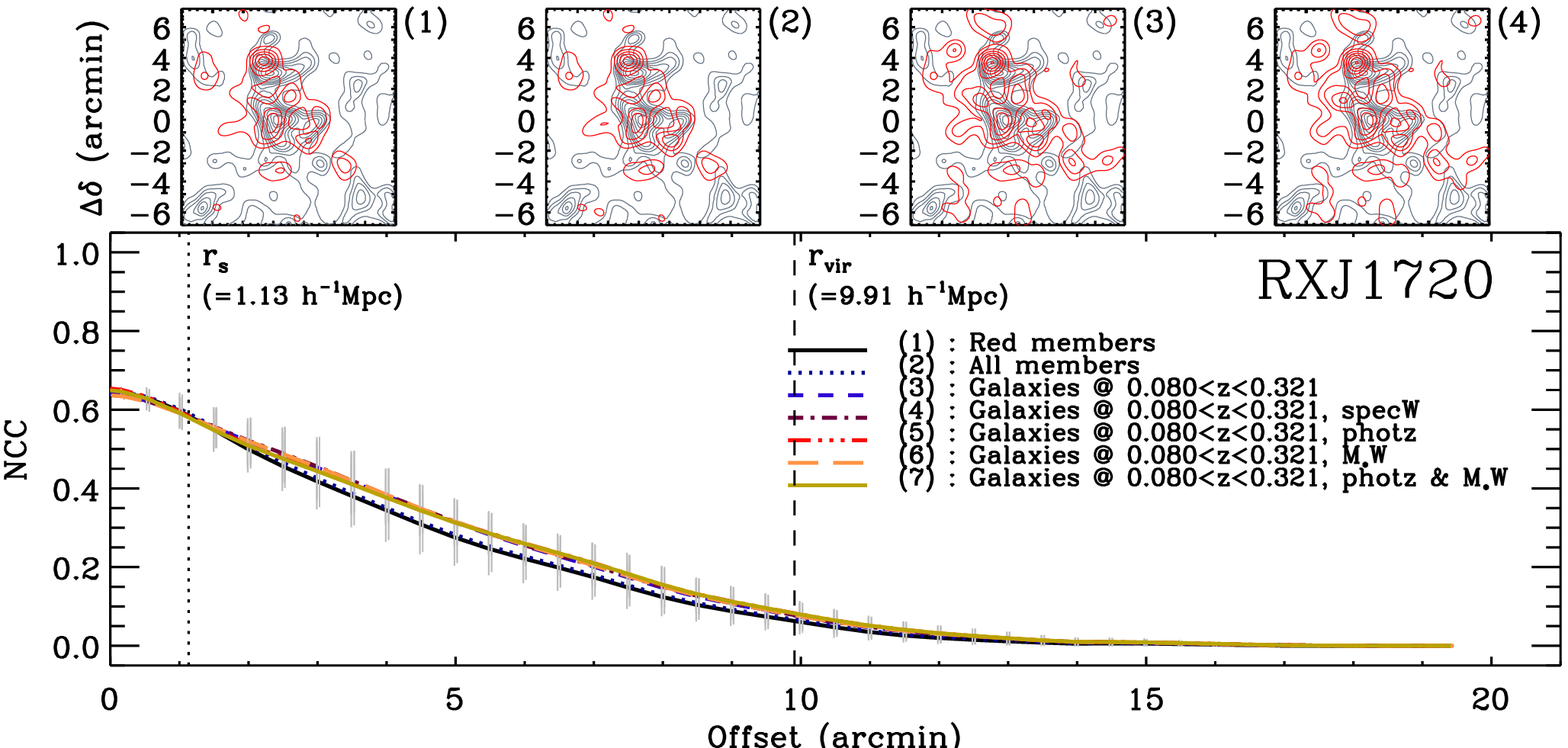} \\
\includegraphics[width=0.8\textwidth]{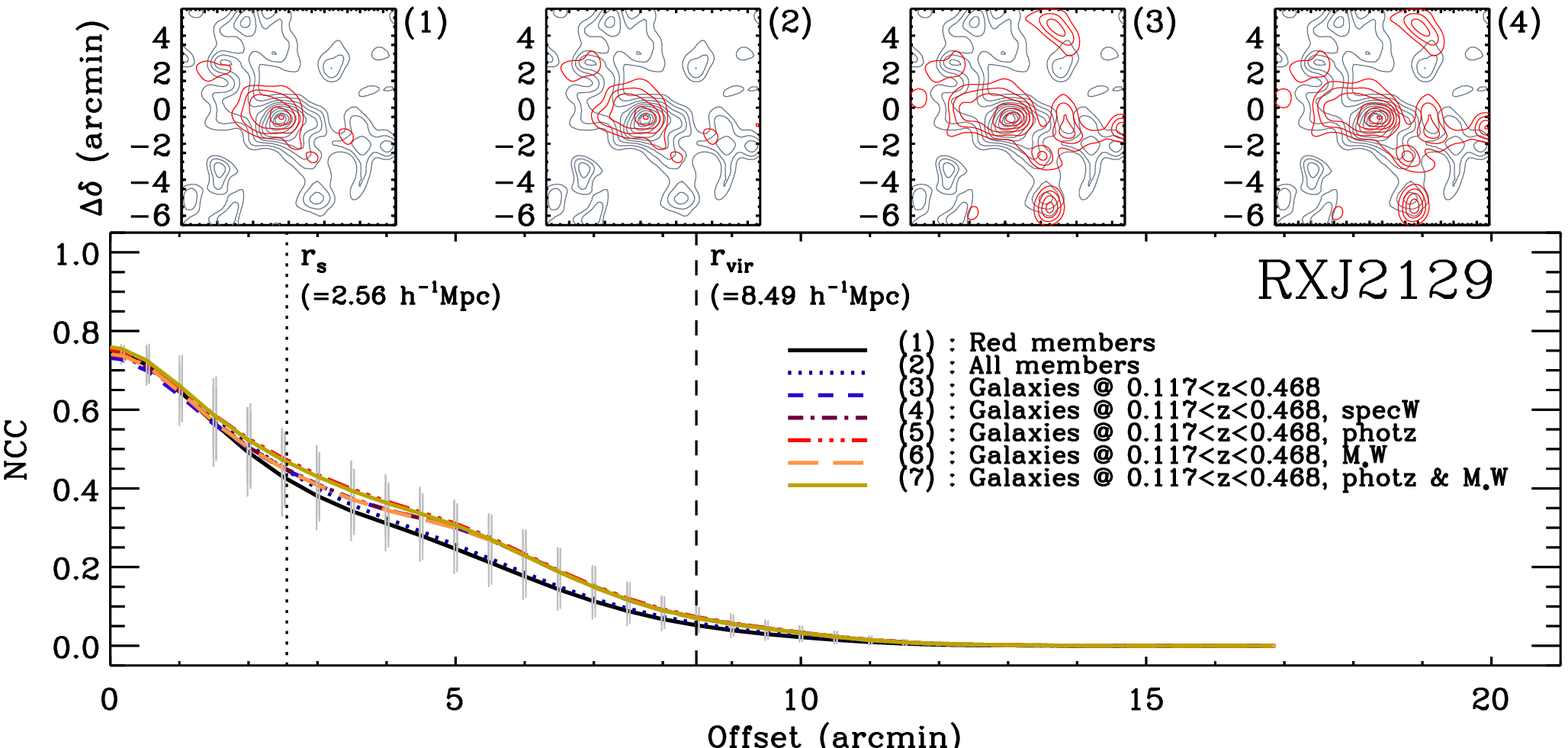} \\
\end{tabular}
\caption{Same as Fig. \ref{fig-cfcc} (except for 2D NCC map),
 but for RXJ1720.1+2638 and RXJ2129.6+0005.
Gray and red contours indicate weak-lensing and 
  galaxy number density maps, respectively.
}\label{fig-cfcc5}
\end{figure*}

Here we discuss the cross correlation 
  results for the other eight clusters.
The plots for the eight clusters are
 in Figures \ref{fig-cfcc2}--\ref{fig-cfcc5} 
 (see Fig. \ref{fig-cfcc} for A383).
 
{\it A267.} The peak of the galaxy number density map 
  based on members alone (top left panels)
  seems to be offset by $\sim$2$^\prime$ from 
  the central peak of the lensing map, but the two peaks
  coincide when we include foreground and background galaxies 
  in the number density map.
The number density maps based on members alone
  show an overdensity only in the central region.
When we include foreground and background galaxies in the map,
  several weak-lensing peaks other than the cluster
  appear in the number density map.
Thus the normalized cross correlation signal
  including foreground and background galaxies
  exceeds those based on cluster members alone at zero offset.
This excess can indicate a non-negligible contribution
  of foreground and background galaxies to the weak-lensing signal,
  but \citet{oka10} found nothing unusual
  in deriving the weak-lensing mass profile for this cluster.
  
{\it A611.} This cluster is the most distant cluster ($z=0.288$) in the sample.
The central region of the weak-lensing map shows a
  north/east-south/west elongation.
The number density contours based on member galaxies alone weakly follow
  this elongation. 
However, the elongated contours are more similar
  to the weak-lensing contours
  when we include foreground and background galaxies.
In other words, foreground and background galaxies
  can affect the apparent ellipticity of
  cluster mass distribution in a weak-lensing map.
The mass profile of this cluster derived 
  from the two-dimensional weak-lensing map
  is correspondingly very noisy (see Fig. 29 in \citealt{oka10}).
The noisy mass profile may result from the complex contribution
  of foreground and background galaxies to cluster shape.

{\it A689.} The X-ray emission of this cluster is dominated
  by a central BL Lac \citep{gil12}, thus
  an X-ray luminosity of this cluster
  is smaller than the other clusters 
  in the sample (see \citealt{rines13} for details).
The weak-lensing map for this cluster
  shows several significant peaks.
The galaxy number density maps based on members alone also show
  peaks in the central and upper regions of the map.
The peaks in the lower regions of the map appear significantly enhanced
  when we include foreground and background galaxies.
The cross correlation signal when we 
   include foreground and background galaxies
   is systematically larger than 
   the one based on members only.
In fact, \citet{oka10} could not derive the mass profile of this cluster
  because the projected mass distribution 
  in the weak-lensing map is so complex.
The difficulty in deriving the cluster mass profile
  occurs in part because the cluster itself is complex and
  the contribution of foreground and background galaxies
  further complicates the weak-lensing signal.
  
{\it A697.} The spectroscopic completeness for this cluster
  within the weak-lensing map
  is one of the highest (89\%) in our sample.
The cluster galaxies are centrally concentrated.
They dominate the signal in the galaxy number density map
  even when we include foreground and background galaxies. 
Thus the radial profiles of the correlation signal
  for several subsamples do not differ significantly.
\citet{oka10} could not fit the mass profile of this cluster with 
  an singular isothermal sphere (SIS) model, 
  but could fit with either of cored isothermal sphere (CIS) or NFW models.

{\it A750.} This cluster is the closest cluster ($z=0.164$) in the sample.
The difference in correlation signal at zero offset
  between cases based on members alone and including foreground/background
  galaxies is the largest in the sample.
This result occurs because
  there are two clusters with similar redshifts aligned
  nearly along the line of sight in this field:
  $z=0.164$ for A750 and $z=0.1767$ for MS 0906.5+1110 
  \citep{car96,rines13,gel13}.
MS 0906.5+1110 is a more X-ray luminous than A750.
Two clusters are often used without a clear distinction between the two.
Their velocity difference is 3250 km s$^{-1}$ in the cluster rest frame.
Our redshift survey resolves these two.
The galaxies in MS 0906.5+1110 are not
  selected as A750 members,
  but contribute significantly to the lensing signal
  (see Section 5 of \citealt{gel13} for details).  
Therefore, the mass based on weak lensing is roughly double the 
  true cluster mass \citep{gel13}.
In fact, \citet{oka10} could not derive a mass profile of this cluster
  because the mass distribution in the weak-lensing map is so complex.

{\it A963.} At small offsets, the correlation signal 
  when we include photometric redshifts for
  foreground and background galaxies are slightly smaller than 
  the one based on members only.
The large dispersion in the number density map
  based on foreground and background galaxies with photometric redshifts
  affects the normalization of the
  correlation signal (see eq \ref{eq-ncc}).
Considering the small difference and the large uncertainty
  in photometric redshifts, the effect is not statistically significant.
Interestingly, 
  \citet{oka10} could not obtain an acceptable fit to the mass profile of 
  this cluster with any of three models (NFW, SIS and CIS). 
This cluster is an exceptional case
  where weak lensing does not provide a robust mass profile
  even though the radial profiles of the cross correlation signal
  for several subsamples do not differ significantly.

{\it RXJ1720.1+2638.} The overall spectroscopic completeness for this cluster
  in the weak-lensing map
  is one of the highest (89\%) in the sample.
The {\it Chandra} X-ray observations 
  show cold fronts in this cluster \citep{maz01},
  probably resulting from sloshing 
  induced by the gravitational perturbation through
  minor mergers with small groups \citep{owe11}.
A secondary peak to the north of the main concentration
  in the weak-lensing map
  appears as an overdensity in the number density map
  based on cluster members alone.  
The one-dimensional cross correlation signals are similar to one another,
  suggesting that the contribution of foreground and background
  galaxies to the weak-lensing map is not significant 
  in this cluster field.
\citet{oka10} found a good fit to their models for
  a weak-lensing mass profile for this cluster.
  
{\it RXJ2129.6+0005.} This cluster has the smallest number
  of galaxies with spectroscopic redshifts
  in the weak-lensing map (N: 156).
The secondary peaks to the west and to the south-southwest
  of the main concentration
  appear only when we include foreground and background galaxies,
  but the amplitudes of the peaks are low.
Therefore,  
  the one-dimensional correlation signal remains similar.
\citet{oka10} found a robust fit to the weak-lensing 
  mass profile for this cluster.

\begin{figure*}
\center
\includegraphics[width=0.8\textwidth]{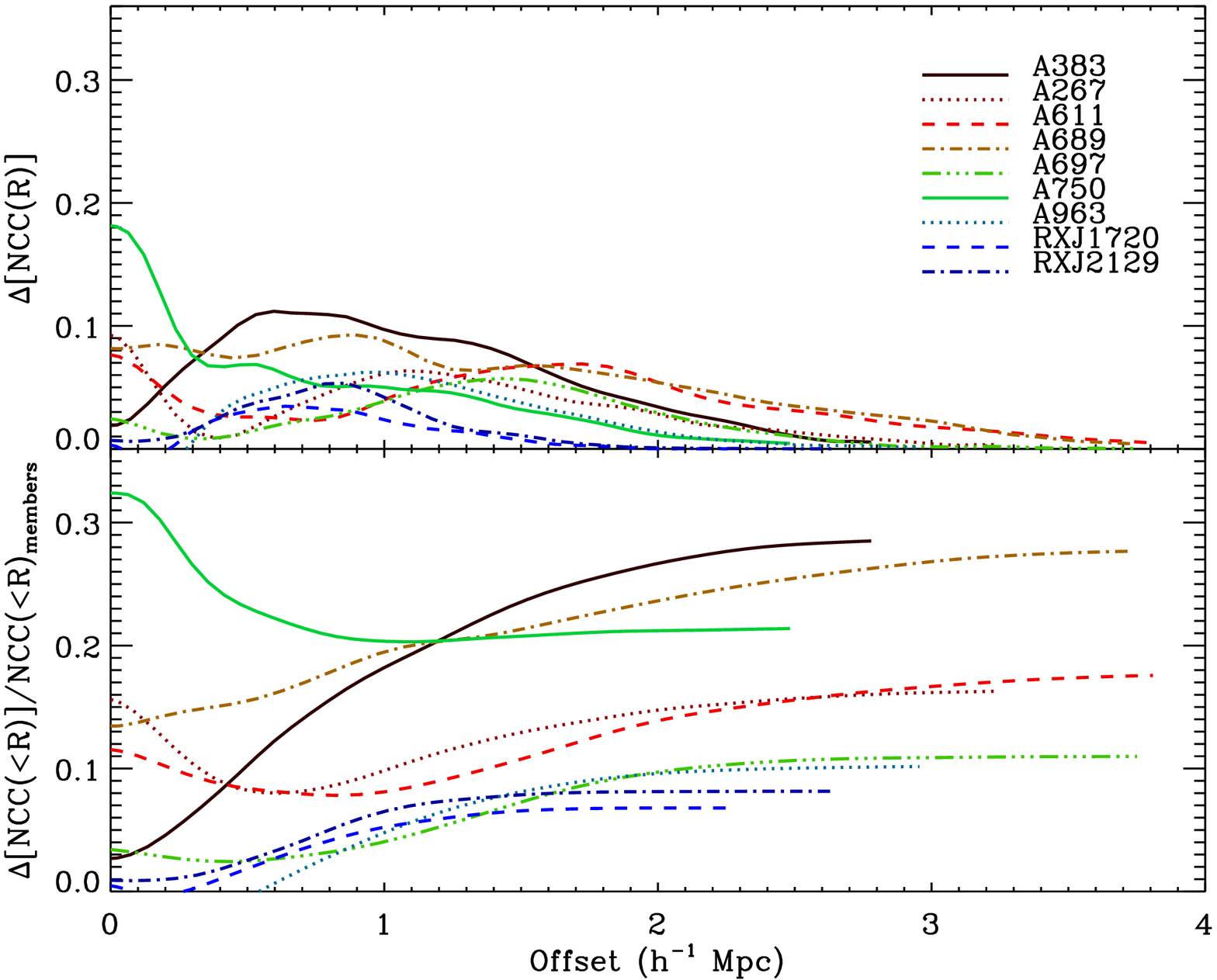}
\caption{({\it Top}) Difference 
  in normalized correlation signal
  for cases with cluster member galaxies 
  (i.e., case 2 in Fig. \ref{fig-cfcc})
  and with all the galaxies in the redshift range 
  0.5$z_{\rm cl}<z<2z_{\rm cl}$, complemented by photometric redshifts
  and weighed by stellar masses (i.e., case 7 in Fig. \ref{fig-cfcc})
  as a function of offset
  for nine clusters in this study.
({\it Bottom}) Same as top panel, but 
 for the cumulative fractional difference in normalized correlation signal.
}\label{fig-fratio}
\end{figure*}

\begin{figure*}
\center
\includegraphics[width=0.7\textwidth]{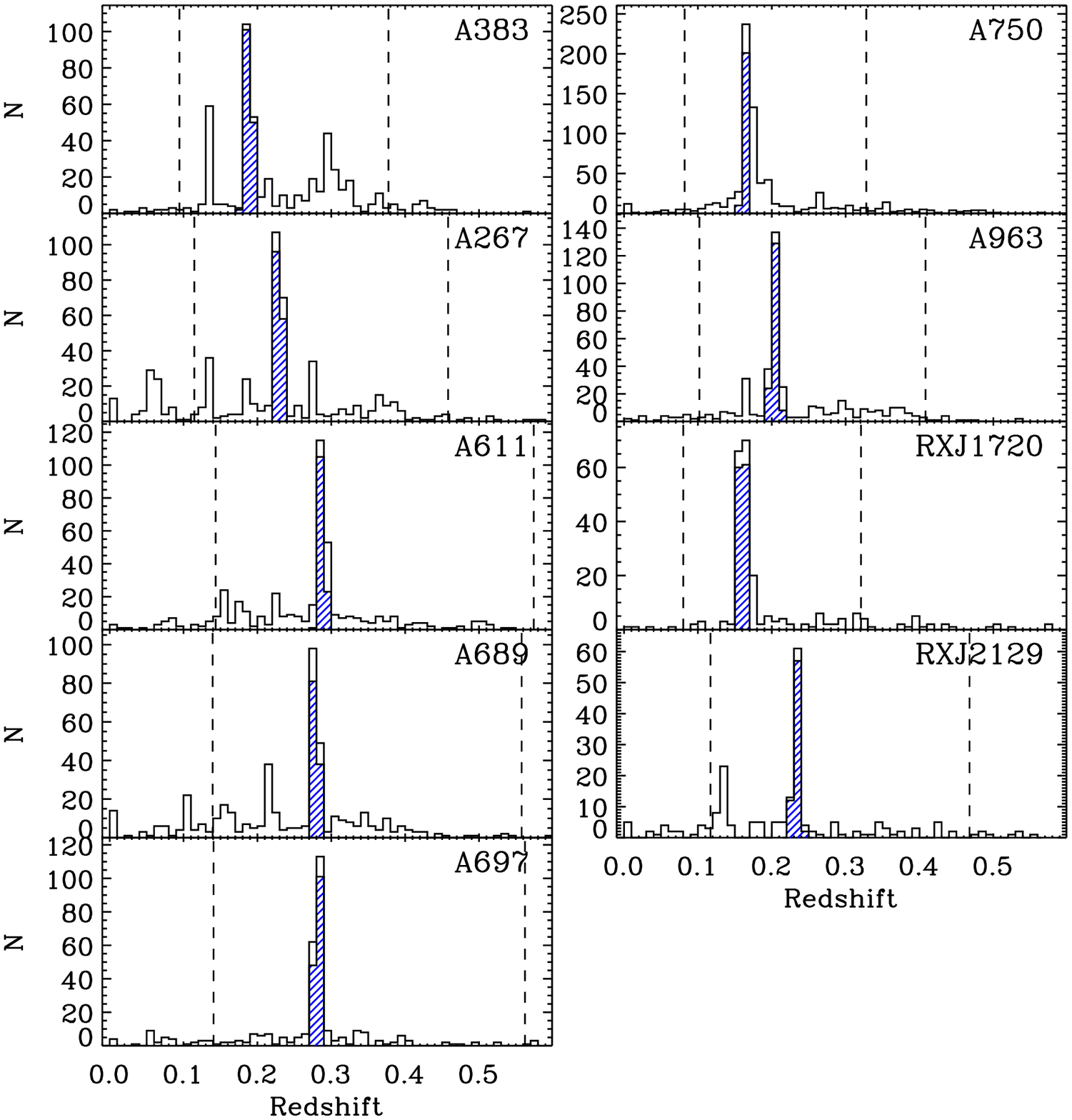}
\caption{Redshift histograms for the galaxies in the Subaru FOVs
  of galaxy clusters.
Blue hatched histogram shows the cluster members.
Open histograms show all the galaxies
  with spectroscopic or photometric redshifts 
  at $m_{\rm r,Petro,0}<20.5$.
Two vertical dashed lines indicate
  the redshift range 
  when we include foreground and background galaxies 
  (i.e., 0.5$z_{\rm cl}$ and 2$z_{\rm cl}$).
}\label{fig-zhist}
\end{figure*}

Comparison of the cross correlation results
  with the weak-lensing mass profiles 
  in \citet{oka10} for the nine clusters
  yields diverse results.
There are three clusters (A689, A750 and A383) 
  where the normalized cross correlation signal
  including foreground and background galaxies
  significantly exceeds the signal based on cluster members alone;
  \citet{oka10} could not derive a stable mass profile for 
  two clusters (e.g., A689 and A750)
  because of the complex projected mass distribution 
  in the weak-lensing map, and
  could not obtain an acceptable fit to the mass profile of A383
  with an NFW, SIS or CIS model.
For two clusters where the normalized cross correlation signal
  including foreground and background galaxies
  slightly exceeds the signal based on cluster members alone at zero offset
  (e.g., A267, A611), \citet{oka10} obtained a robust weak-lensing mass profile.
\citet{oka10} derived a stable weak-lensing mass profile
  for two clusters where there is no significant difference between
  the normalized cross correlation signal for several galaxy subsamples
  (e.g., RXJ1720.1+2638, RXJ2129.6+0005).
\citet{oka10} could not obtain an acceptable model fit
  to the derived mass profiles of two clusters  (e.g., A697, A963)
  even though there is no significant difference in
  the normalized cross correlation signal for the galaxy subsamples
  we investigate here.
Although the nine cluster show diverse results,
  a normalized cross correlation signal
  including foreground and background galaxies
  that significantly exceeds the signal based on cluster members alone
  appears to be a good proxy for an underlying systematic problem 
  in the interpretation of the projected mass distribution as revealed
  by the weak-lensing map. 
Problems may result from resolved substructure within the cluster (e.g., A689) 
  and/or from the contribution of large-scale structure superimposed 
  along the line of sight (e.g., A383, A750).

\section{DISCUSSION}\label{discuss}

We use dense redshift surveys in the fields of nine $z\sim0.2$ galaxy clusters
  to cross correlate galaxy number density maps 
  with weak-lensing maps.
The cross correlation signal
  when we include foreground and background galaxies
  exceeds the one based on cluster members alone 
  (see Figs \ref{fig-cfcc}--\ref{fig-cfcc5});
  the cross correlation for the full sample is
  not negligible even outside the cluster virial radii.
  
We summarize the results of the cross correlation 
  for the nine clusters in Figure \ref{fig-fratio}.
We plot the difference in the normalized correlation signal between
  the cases including all the galaxies around clusters and 
  based on cluster members alone (i.e., case 7 and 2 in Fig. \ref{fig-cfcc}) 
  in the top panel.
The bottom panel shows the cumulative ($<$R), 
  fractional excess for the case
  including all the galaxies around the cluster
  relative to the case based on cluster members alone
  (i.e., case 7 and 2 in Fig. \ref{fig-cfcc}).

The bottom panel shows that
 the cumulative fractional excess changes
 with offset, and increases up to 30\%.
The typical statistical error in the cumulative 
  fractional excess is $3-5\%$.
The nine clusters show different patterns,
  reflecting different large-scale structure
  along the line of sight.
At the typical virial radius of our cluster samples 
  (i.e., $\sim$ 1.3 $h^{-1}$Mpc),
  the cumulative fractional excess is 5--23\%.
These excesses are roughly consistent with the results based on
  simulation data \citep{hoe11};
  uncorrelated large-scale structure 
  along the line of sight contributes to an uncertainty 
  in weak-lensing cluster mass
  of $10-25\%$ for $z\sim0.2$ clusters (see also \citealt{hoe01}
  for analytical prediction).  
The increase of the cumulative fractional excess with offset
  is also consistent with the idea that
  the weak-lensing analysis tends to overestimate cluster masses
  in the outer regions (e.g., see Fig. 13 in \citealt{gel13}).
Studies of much larger cluster samples suggest 
  that one can determine robust 3D mass profiles of galaxy clusters 
  from weak lensing measurements based on the ensemble;
  a halo model approach to analysis of the ensemble 
  can treat the correlated structure  \citep{john07a,john07b,lea10}.
 
One interesting aspect of Figure \ref{fig-fratio} is that
  the fractional excess is significant even at zero offset
  for some clusters (e.g., A267, A611, A689 and A750).
Thus we might expect a significant number of galaxies 
  inside the Subaru FOV that are not cluster members, 
  but that contribute to the lensing signal.
Figure \ref{fig-zhist} shows a redshift histogram
  for the galaxies in the weak-lensing map of each cluster.
The hatched and open histograms show the cluster members
  and all the galaxies with spectroscopic/photometric 
  redshifts at $m_{\rm r,Petro,0}\leq20.5$, respectively.
The histogram clearly shows that the fields of the clusters with large
  fractional excess have
  non-member galaxies close to the mean cluster redshift.
  
Based on simulations, 
  \citet{bk11} emphasize
  that large-scale structure associated (or correlated) with 
  individual clusters
  makes a non-negligible contribution ($\sim20\%$)
  to the scatter of weak-lensing mass estimates
  in addition to the contribution of 
  uncorrelated large-scale structure along the line of sight 
  ($10-25\%$, \citealt{hoe11}).
We highlight the field of A750 in Section \ref{comments}
  where the contribution of the lensing signal from 
  a nearby superimposed cluster is about a factor of two.

The vertical dashed lines in Figure \ref{fig-zhist} 
  indicate the redshift range that we use to 
  include foreground and background galaxies
  when we make the galaxy number density map for each cluster
  (i.e., 0.5$z_{\rm cl}$ and 2$z_{\rm cl}$).
The nine clusters are at similar redshifts (i.e., $z\sim$0.2),
  thus the redshift ranges for including foreground and background galaxies
  are also similar.
Moreover, examination of this plot also shows that
  the amplitude of the cumulative fractional excess in the bottom panel
  of Figure \ref{fig-fratio}
  does not depend on the cluster redshift, suggesting that
  the small difference in cluster redshifts in our sample
  does not introduce a bias.

The spectroscopic completeness of our redshift survey for the nine clusters is
  very high (70--89\% at $m_{\rm r,Petro,0}\leq20.5$).
However, our redshift surveys are not deep enough to include
  a large number of galaxies at high redshift end of the lensing kernel
  (e.g., galaxies fainter than $m_{\rm r,Petro,0}=20.5$ 
  at $0.1\lesssim z\lesssim0.4$ for $z\sim0.2$ cluster).
Thus our estimates of the contribution of
  large-scale structure along the line of sight to the cluster lensing maps
  could be lower limits;
  the real contribution could be larger even though
  photometric redshifts mitigate this issue to some extent.

\citet{oka10} studied the mass profiles of the nine clusters in this study
  derived from the two-dimensional weak-lensing maps.
They used three mass models to fit the derived mass profiles: 
  NFW, SIS and CIS.
Interestingly, they could not derive the mass profiles of A750 and A689 
  because the mass distribution in the weak-lensing maps 
  is complex in these systems.
They could fit the mass profiles with their three models 
  for most clusters,
  but could not obtain an acceptable fit to the mass profile of 
  A383 with any model.
These three clusters (A383, A689 and A750)
  show the largest fractional excesses ($11-30\%$) in 
  the integrated normalized correlation signal
  at offset $>$0.5 $h^{-1}$ Mpc.
The range of possible impacts of the superimposed structure
  on cluster mass estimates is large: a few percent for A383 \citep{gel14}, 
  and up to a factor of two for A750 \citep{gel13}.
These results suggest that
  the excess in the integrated cross correlation signal
  could be a useful proxy for assessing the reliability of 
  weak-lensing cluster mass estimates.

We cross correlate two independent measurements, 
  the galaxy number density map and 
  the lensing convergence field, $\kappa$ for each cluster. 
As suggested by \citet{van13}, it would be interesting to use 
  the galaxy number density map to reconstruct a model convergence map 
  for comparison with the observed one. 
However, this procedure requires knowledge positions and redshifts 
  of the source galaxies that are not currently available.

\section{CONCLUSIONS}\label{sum}
  
We use dense redshift surveys in the fields of nine $z\sim0.2$ galaxy clusters
  to compare the structures identified 
  in weak-lensing and galaxy number density maps.
We combine 2087 new MMT/Hectospec redshifts and 
  the data in the literature to make
  the overall spectroscopic completeness 
  in the weak-lensing maps high (70--89\%) and uniform.
With these dense redshift surveys,
  we first construct galaxy number density maps using several galaxy subsamples.
Our primary results are:

\begin{enumerate}

\item The global morphology of the spatial distribution
  of cluster members alone is similar to the shape of the cluster peak
  in weak-lensing maps.
However, in some clusters (e.g., A611),
  the apparent shape of the cluster peak
  in the weak-lensing map may be affected
  by the contribution of foreground and background galaxies.

\item The red cluster galaxies dominate 
  the cluster weak-lensing signal, and
  blue cluster galaxies contribute little to
  the number density maps of cluster members.
These results suggest that
  the red populations are reliable tracers
  of the mass distribution of galaxy clusters
  provided that the selection is dense and broad enough 
  around the red sequence.  
This result supports the approach often taken 
  in strong lensing analysis \citep{bro05,zit09,med10}.

\item The correspondence between the galaxy number density 
  and weak-lensing maps
  is even more remarkable when we include foreground and background galaxies
  in the galaxy number density maps,
  reflecting the contribution of superimposed large-scale structure
  along the line of sight to the $\kappa$ map.
\end{enumerate}

We cross correlate the galaxy number density maps
  with the weak-lensing maps, and we find:

\begin{enumerate}

\item The cross correlation signal
  when we include foreground and background galaxies
  at 0.5$z_{\rm cl}<z<2z_{\rm cl}$
  is always larger than for the case with cluster members alone.
The fractional excess of the integrated normalized correlation signal
  for the case including foreground and background galaxies
  relative to the case base on cluster members alone
  is $10-23$\% at the cluster virial radius. 
This excess can be as high as 30\% depending on the cluster.

\item Superimposed structure close to the cluster
  in redshift space
  contributes to the weak-lensing peaks more significantly
  than unrelated large-scale structure along the line of sight.
  
\item The mass profiles of three clusters (A383, A689 and A750)
  with the largest fractional excesses ($11-30\%$) of 
  the integrated normalized correlation signal
  are also not well constrained in weak lensing \citep{oka10}.
Thus the excess in 
  the integrated normalized correlation signal
  could be a useful proxy for assessing the reliability of 
  weak-lensing cluster mass estimates.

\end{enumerate}

A dense redshift survey
  of galaxy clusters is important for understanding the meaning
  of a weak-lensing $\kappa$ map.
We plan to extend this study to a larger cluster sample 
  including merging clusters
  in a forthcoming paper (H. S. Hwang et al., in preparation).

Future exploration with deep spectroscopic or even photometric redshift
  surveys covering all the galaxies including faint ones
  in the lensing kernel is important for refining the estimates
  of the contribution
  of large-scale structure along the line of sight accurately.
To resolve the structure near clusters that contribute 
  to weak-lensing maps significantly, 
  a spectroscopic survey is crucial.
Techniques that treat the redshift survey simultaneously
  with weak lensing
  may eventually provide very powerful probes of mass distribution
  uncontaminated by superimposed structure along the line of sight.

{\it Facility:} \facility{MMT}

\acknowledgments

We thank the anonymous referee for helpful comments.
We thank Perry Berlind and Michael Calkins for operating the Hectospec,
  and Susan Tokarz and Sean Moran for reducing the Hectospec data. 
We also thank Dan Coe, Ian Dell'Antonio, Nobuhiro Okabe, 
  Graham Smith, Keiichi Umetsu and Adi Zitrin
  for helpful comments in early stages of this work.
The Smithsonian Institution supports the research of MJG and HSH.
AD acknowledges partial support from the INFN grant Indark and from the grant
Progetti di Ateneo $/$ CSP TO$\_$Call2$\_$2012$\_$0011 ``Marco Polo''
of the University of Torino.
Funding for SDSS-III has been provided by the Alfred P. Sloan Foundation, the Participating Institutions, the National Science Foundation, and the U.S. Department of Energy Office of Science. The SDSS-III web site is http://www.sdss3.org/.
SDSS-III is managed by the Astrophysical Research Consortium for the Participating Institutions of the SDSS-III Collaboration including the University of Arizona, the Brazilian Participation Group, Brookhaven National Laboratory, Carnegie Mellon University, University of Florida, the French Participation Group, the German Participation Group, Harvard University, the Instituto de Astrofisica de Canarias, the Michigan State/Notre Dame/JINA Participation Group, Johns Hopkins University, Lawrence Berkeley National Laboratory, Max Planck Institute for Astrophysics, Max Planck Institute for Extraterrestrial Physics, New Mexico State University, New York University, Ohio State University, Pennsylvania State University, University of Portsmouth, Princeton University, the Spanish Participation Group, University of Tokyo, University of Utah, Vanderbilt University, University of Virginia, University of Washington, and Yale University.
This research has made use of the NASA/IPAC Extragalactic Database (NED) which is operated by the Jet Propulsion Laboratory, California Institute of Technology, under contract with the National Aeronautics and Space Administration.

\bibliographystyle{apj} 
\bibliography{ref_hshwang} 

\appendix

\section{Appendix Material}

\begin{figure*}
\center
\begin{tabular}{c}
\includegraphics[width=0.8\textwidth]{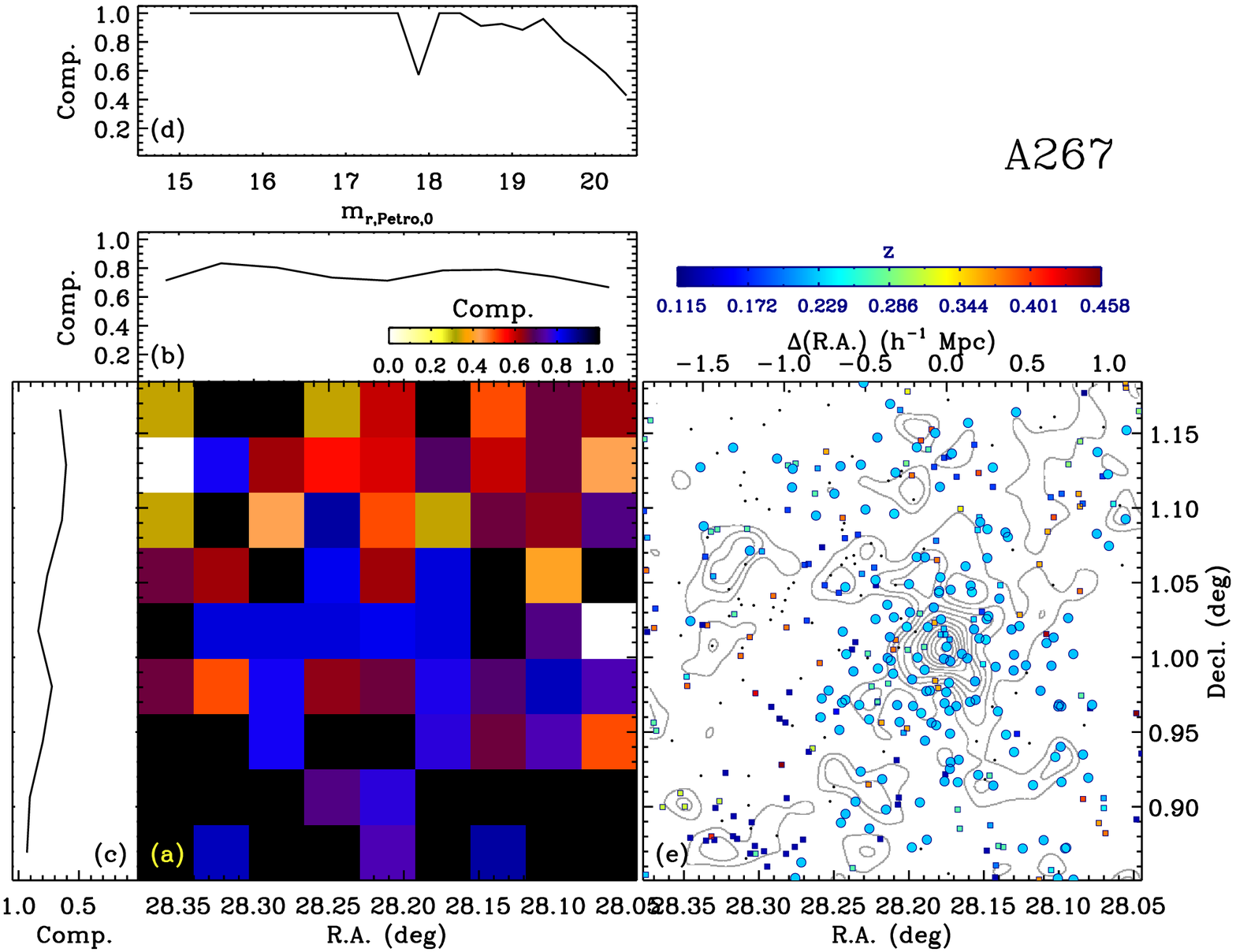} \\
\includegraphics[width=0.8\textwidth]{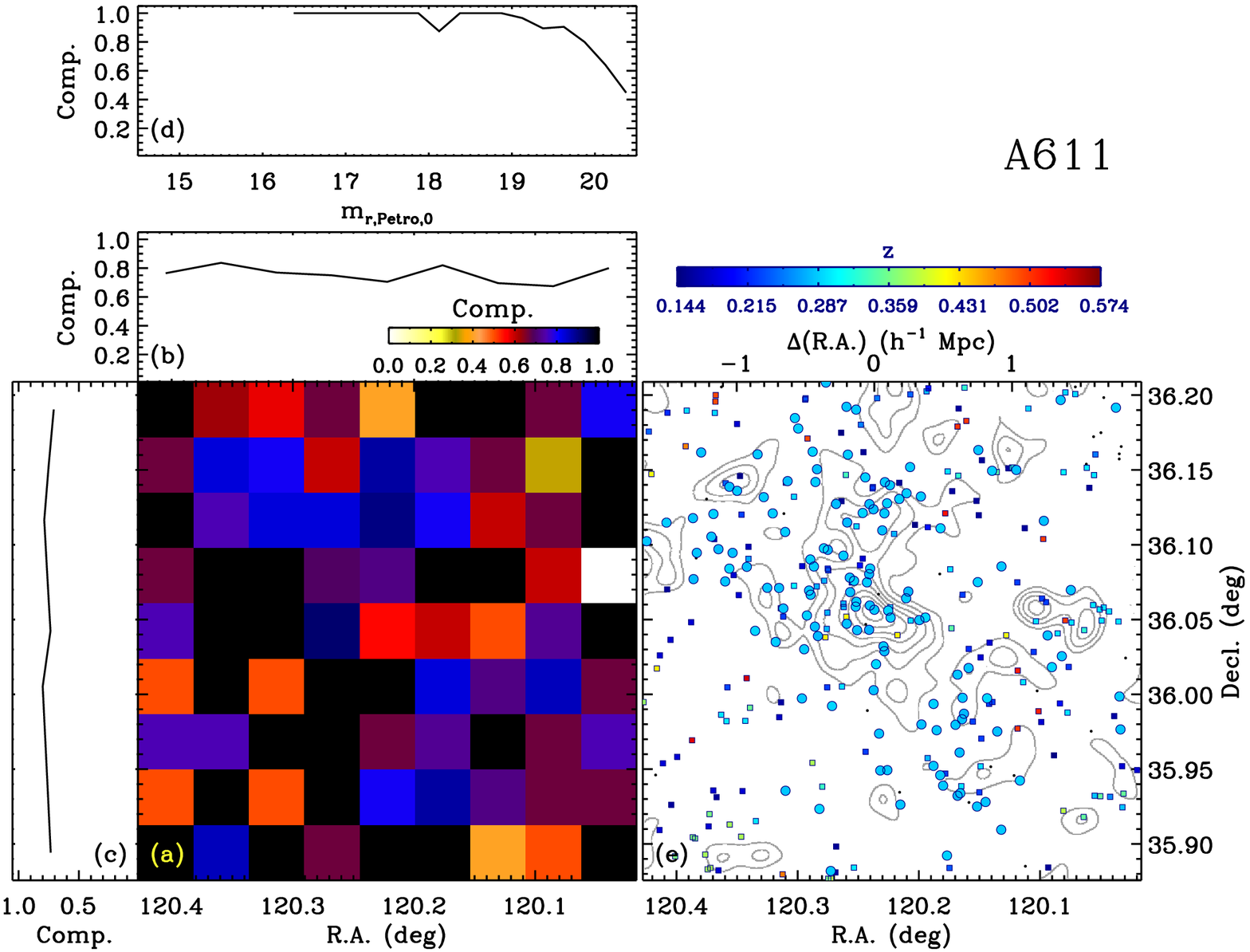}
\end{tabular}
\caption{Same as Fig. \ref{fig-a383spat},
 but for A267 and A611.
}\label{fig-spat2}
\end{figure*}

\begin{figure*}
\center
\begin{tabular}{c}
\includegraphics[width=0.8\textwidth]{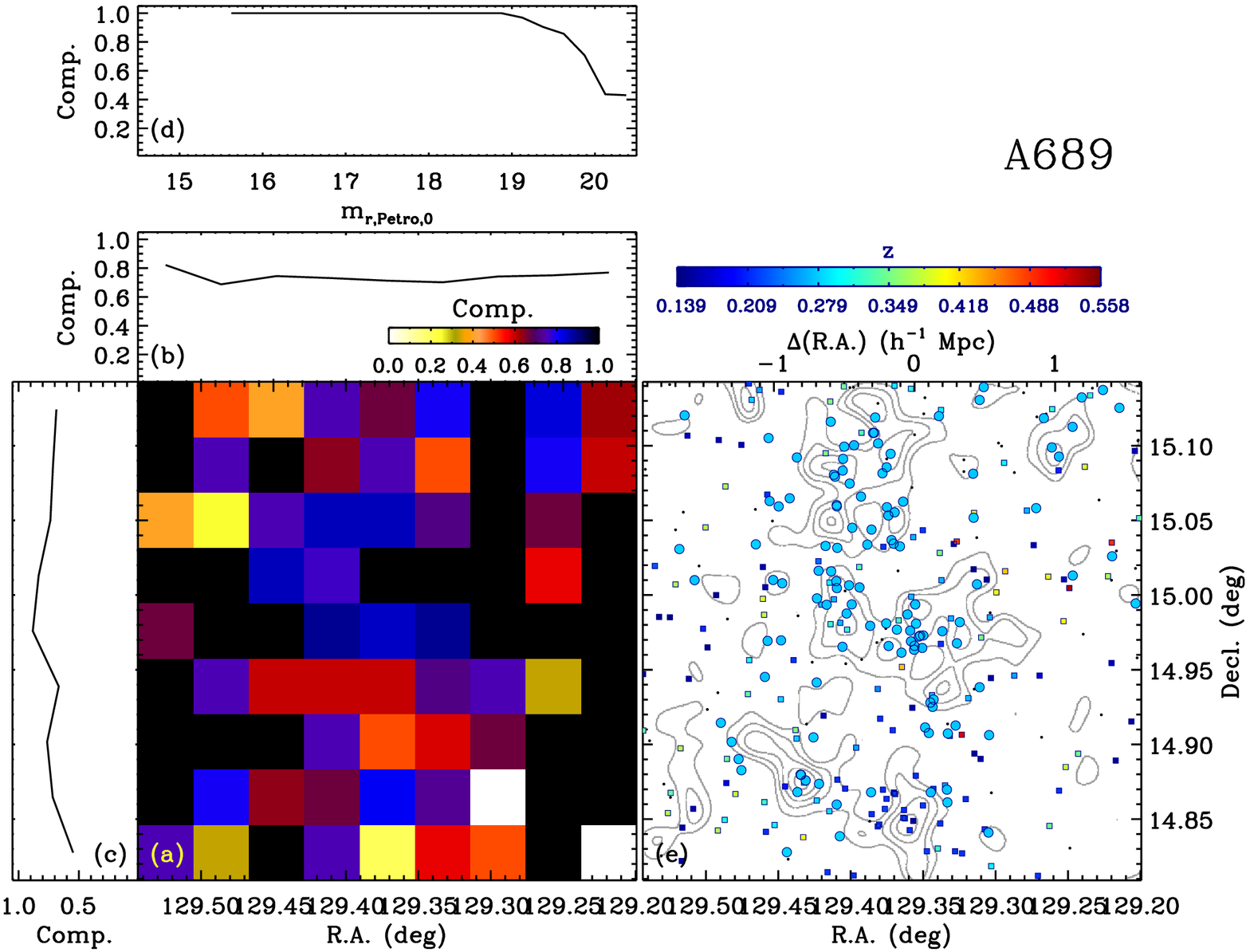} \\
\includegraphics[width=0.8\textwidth]{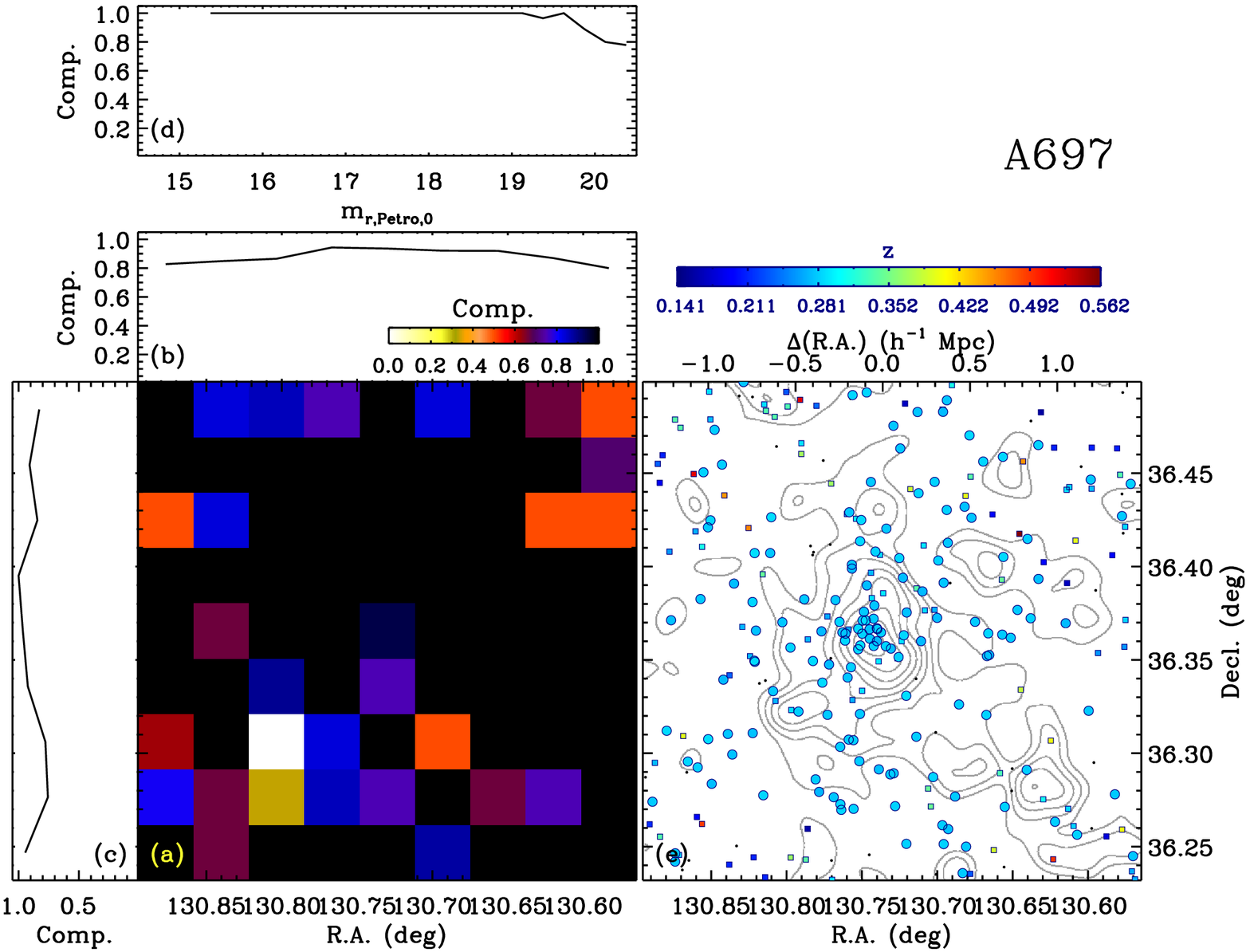}
\end{tabular}
\caption{Same as Fig. \ref{fig-a383spat},
 but for A689 and A697.
}\label{fig-spat3}
\end{figure*}

\begin{figure*}
\center
\begin{tabular}{c}
\includegraphics[width=0.8\textwidth]{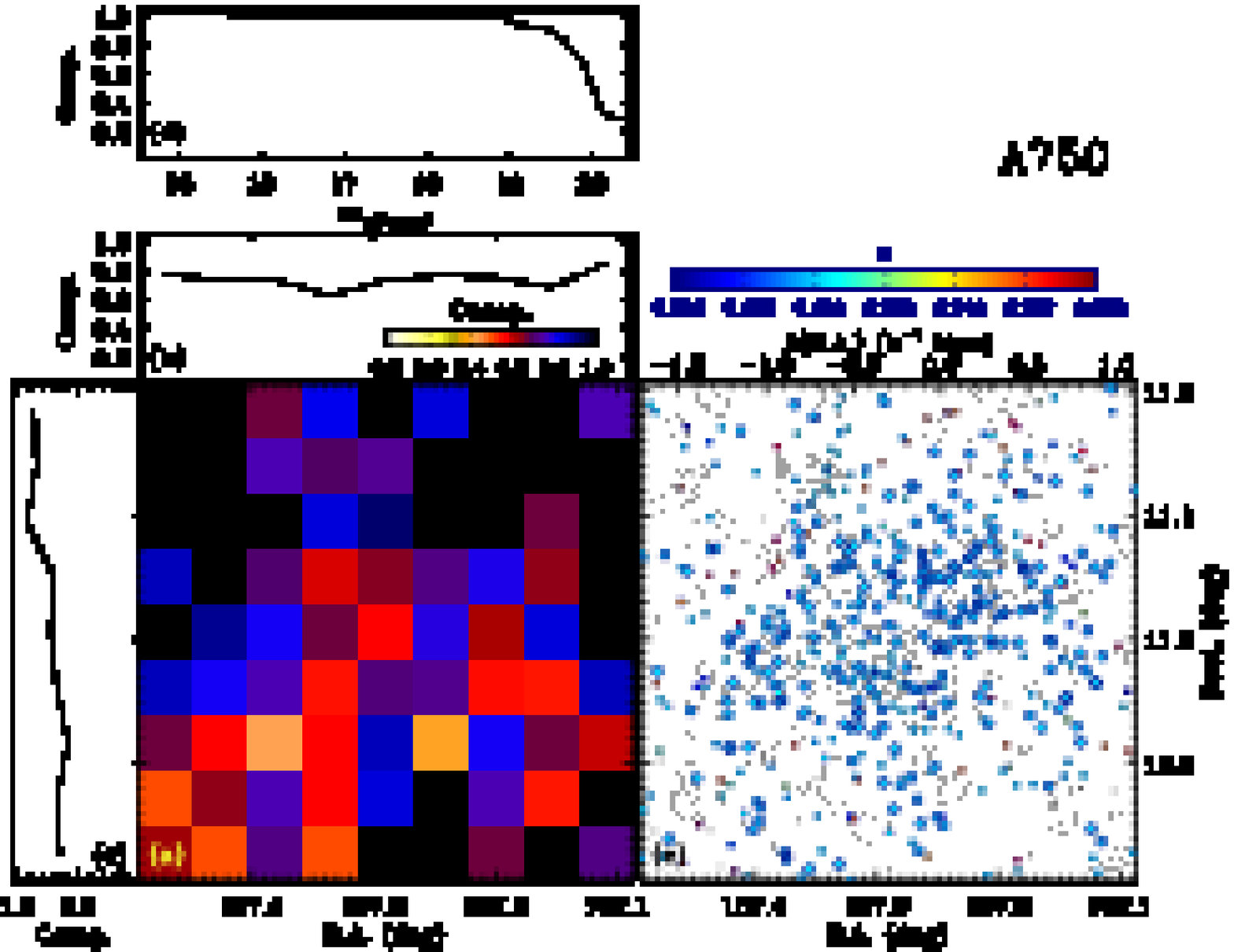} \\
\includegraphics[width=0.8\textwidth]{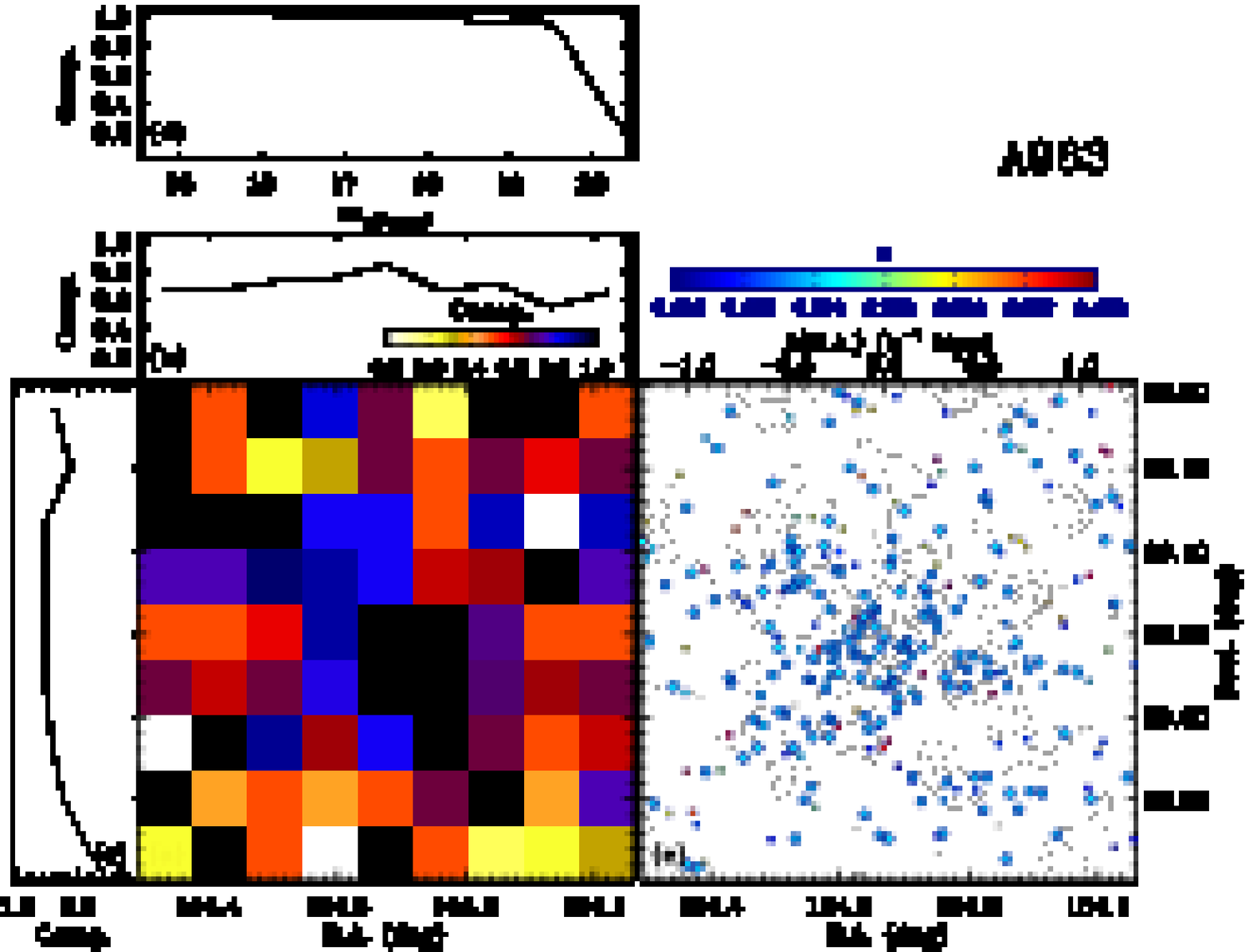}
\end{tabular}
\caption{Same as Fig. \ref{fig-a383spat},
 but for A750 and A963.
}\label{fig-spat4}
\end{figure*}

\begin{figure*}
\center
\begin{tabular}{c}
\includegraphics[width=0.8\textwidth]{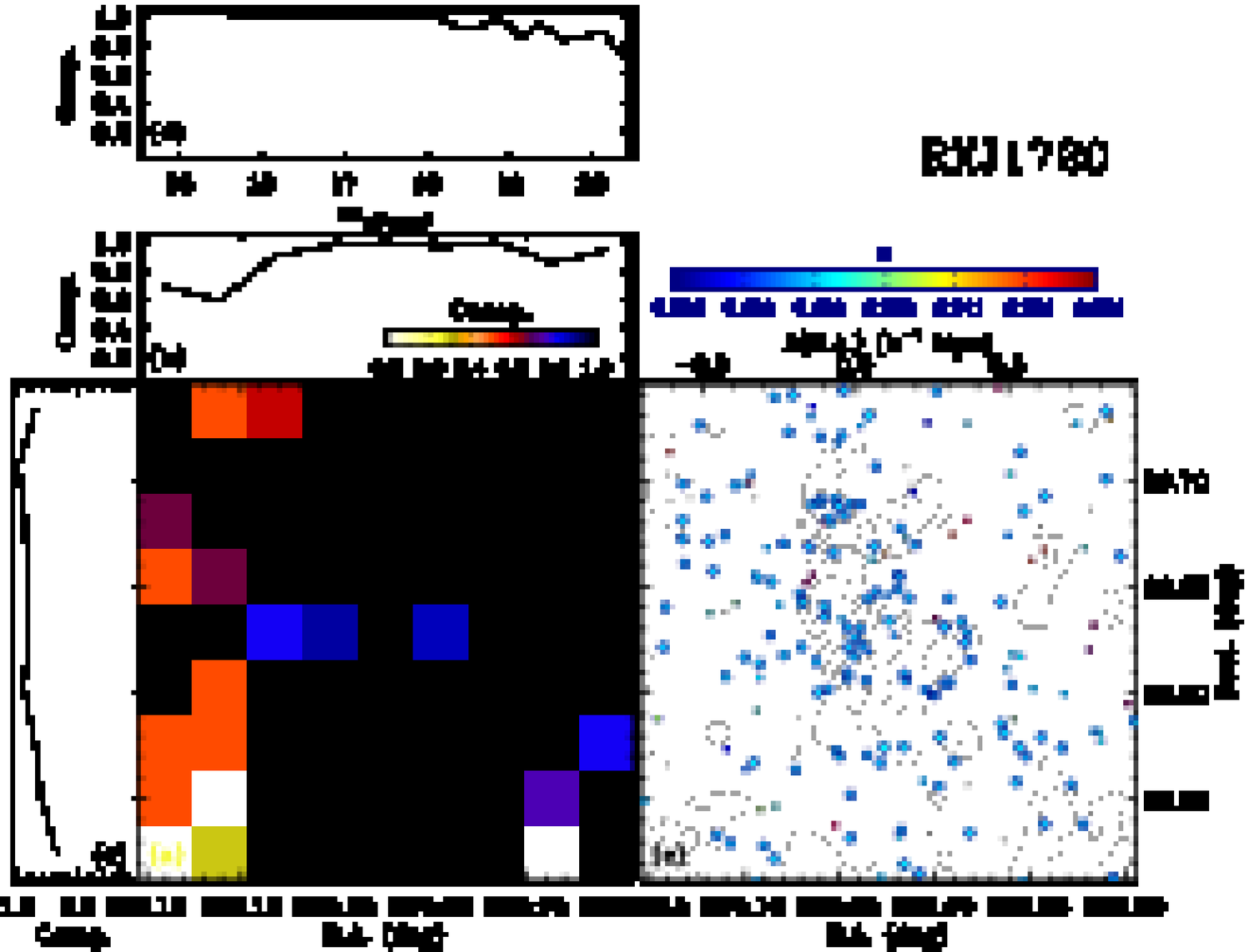} \\
\includegraphics[width=0.8\textwidth]{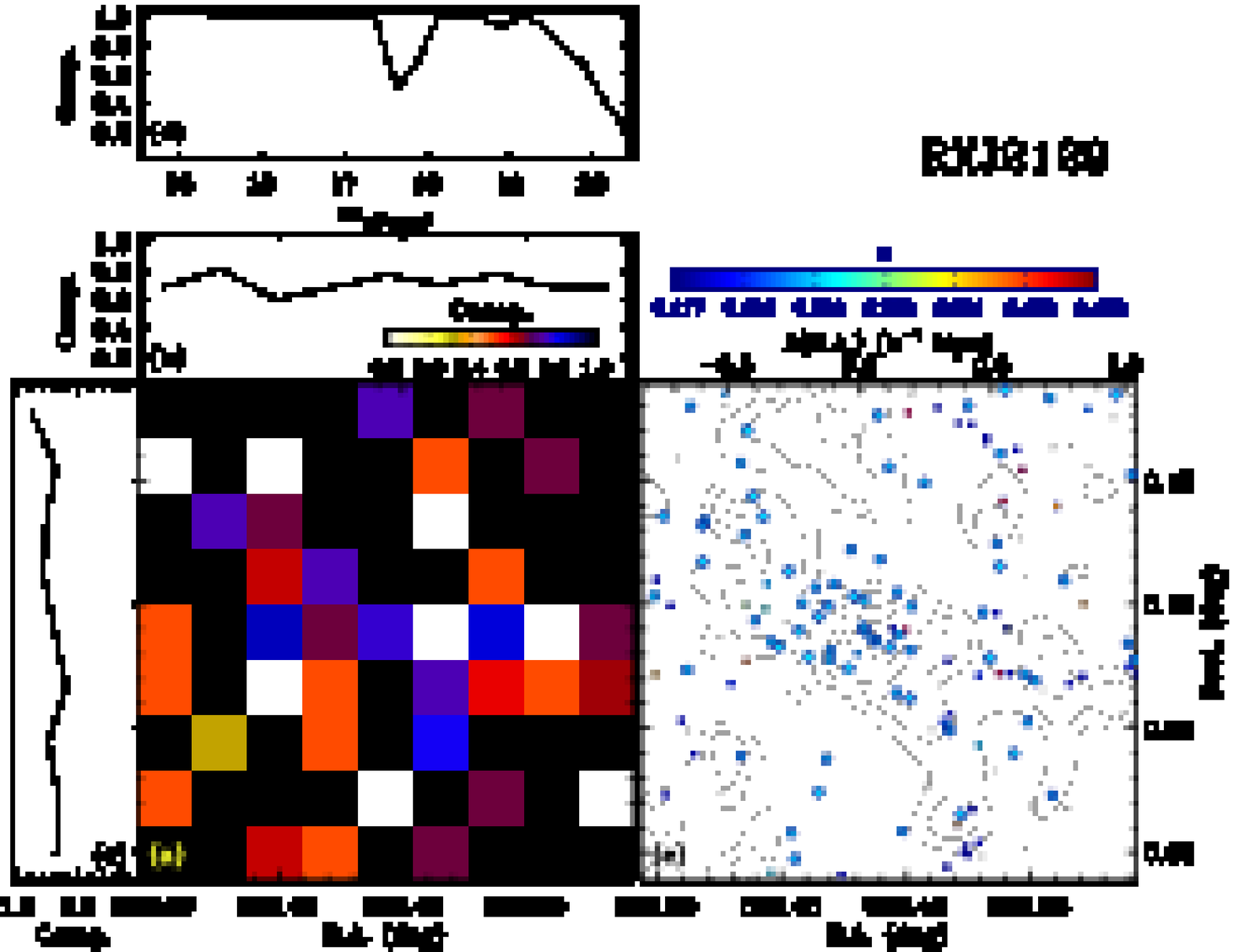}
\end{tabular}
\caption{Same as Fig. \ref{fig-a383spat},
 but for RXJ1720.1+2638 and RXJ2129.6+0005.
}\label{fig-spat5}
\end{figure*}

\begin{figure*}
\center
\includegraphics[width=0.65\textwidth]{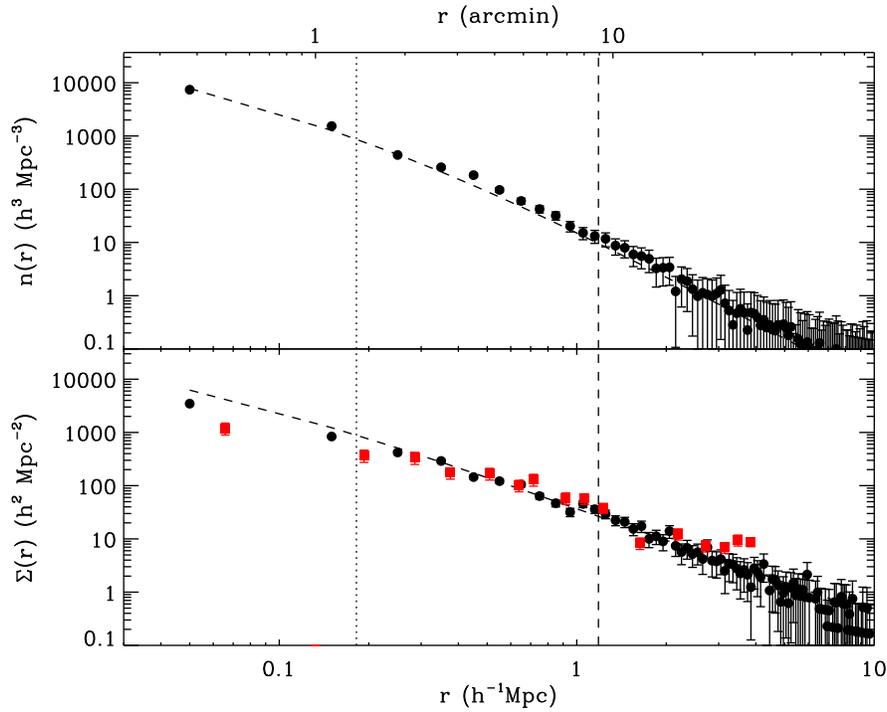}
\caption{({\it Top}) Galaxy number density profile
  for a simulated cluster (filled circles).
Dashed line is the input NFW profile 
  with $r_{200}=1.184$ ($h^{-1}$Mpc) and $c_{200}=6.51$ \citep{new13}.
Vertical dotted and dashed lines indicate
  $r_{s}$ and $r_{200}$, respectively.
({\it Bottom}) Same as top panel, but 
 for projected galaxy number density profile.
Red squares are observed galaxy number density profile of A383
  after completeness correction derived in this study.
}\label{fig-simnum}
\end{figure*}

\begin{figure*}
\center
\includegraphics[width=0.65\textwidth]{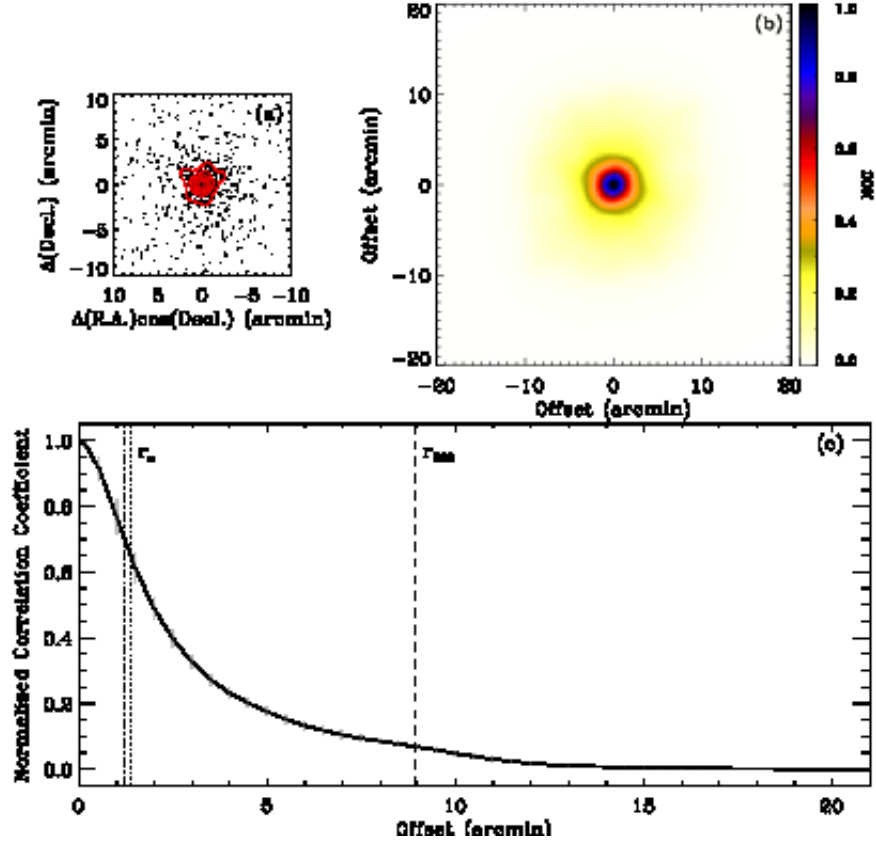}
\caption{Normalized auto correlation of 
  the galaxy number density map for a simulated cluster.
({\it Top left}) Spatial distribution of galaxies (black dots) with
  galaxy number density contours (red).
({\it Top right}) Two-dimensional normalized auto correlation map.
({\it Bottom}) Azimuthally averaged correlation signal
  as a function of offset.
Vertical dot-dashed, dotted and dashed lines indicate
  a smoothing scale FWHM$=1^\prime.2$, $r_{s}$ and $r_{200}$, respectively.
}\label{fig-simcor}
\end{figure*}

\end{document}